\documentclass[appendixfloats]{emulateapj}
\usepackage{hyperref}
\usepackage{threeparttable}
\usepackage{amsmath} 
\usepackage{natbib}
\usepackage{url}
\shortauthors{Marchenko et al.}
\shorttitle{Multiwavelength Structure of 3C\,273 Jet}

\begin{document}

\title{Novel Analysis of the Multiwavelength Structure \\ of Relativistic Jet in Quasar 3C\,273}

\author{Volodymyr Marchenko$^1$, D.E. Harris$^2$, Micha{\l} Ostrowski$^1$, {\L}ukasz Stawarz$^{1}$,\\ Artem Bohdan$^3$, Marek Jamrozy$^1$, and Bohdan Hnatyk$^4$}

\medskip

\affil{$^1$Astronomical Observatory, Jagiellonian University, ul. Orla 171, 30-244 Krak\'ow, Poland}
\affil{$^2$Harvard Smithsonian Center for Astrophysics, 60 Garden St, Cambridge, MA 02138, USA}
\affil{$^3$Institute of Nuclear Physics, Polish Academy of Sciences, ul. Radzikowskiego 152, 31-342 Krak\'ow, Poland}
\affil{$^4$Astronomical Observatory of Kyiv National University, 3 Observatorna Str., 04053 Kyiv, Ukraine}

\medskip

\email{email: {\tt volodymyr.marchenko@oa.uj.edu.pl}}

\label{firstpage}

\begin{abstract}

We present a detailed analysis of the best-quality multi-wavelength data gathered for the large-scale jet in the core-dominated quasar 3C\,273. We analyze all the archival observations of the target with the {\it Chandra} X-ray Observatory, the far-ultraviolet observations with the {\it Hubble} Space Telescope, and the 8.4\,GHz map obtained with the Very Large Array. In our study we focus on investigating the morphology of the outflow at different frequencies, and therefore we apply various techniques for the image deconvolution, paying particular attention to a precise modeling of the {\it Chandra} and {\it Hubble} point spread functions. We find that the prominent brightness enhancements in the X-ray and far-ultraviolet jet of 3C\,273 --- the ``knots'' --- are not point-like, and can be resolved transversely as extended features with sizes of about $\simeq 0.5$\,kpc. Also, the radio outflow is wider than the deconvolved X-ray/ultraviolet jet. We have also found circumstantial evidence that the intensity peaks of the X-ray knots are located systematically upstream of the corresponding radio intensity peaks, with the projected spatial offsets along the jet ranging from $\lesssim 0.2$\,kpc up to $\simeq 1$\,kpc. We discuss our findings in the wider context of multi-component models for the emission and structure of large-scale quasar jets, and speculate on the physical processes enabling an efficient acceleration of the emitting ultra-relativistic electrons along the entire jet length that exceeds 100\,kpc.

\end{abstract}


\section{Introduction} 
\label{S:intro}

Large-scale jets produced in active galactic nuclei (AGN) are spectacular manifestations of an efficient extraction of energy and angular momentum from supermassive black holes and their accretion disks \citep{begelman84}. For many years since their discovery, these structures have been studied almost exclusively at radio frequencies \citep{bridle84}, and only recently the new generation of sensitive, high-resolution optical and X-ray instruments, in particular the {\it Hubble} Space Telescope and the {\it Chandra} X-ray Observatory, enabled a truly multiwavelength investigation \citep[see][for a review]{harris06}. In the cases of distant quasars, an in-depth analysis of the optical or X-ray jet structure is however often hampered by a limited photon statistics; only a few targets are bright enough and sufficiently extended on the sky that a more detailed comparison between radio, optical, and X-ray maps can, in principle, be attempted.

Till now, tens of large-scale jets in various types of AGN have been detected in X-rays, and several related {\it Chandra} surveys and population studies have been presented in the literature \citep{harris02,harris06,sambruna02,sambruna04,marshall05,marshall11,kataoka05,massaro11,hogan11}.\footnote{\url{http://hea-www.harvard.edu/XJET/}} Due to the fact that the X-ray fluxes of various segments of the jets in luminous quasars exceed the extrapolation of radio--to--optical synchrotron continua, this X-ray emission is interpreted as either inverse-Comptonization of the cosmic microwave background (CMB) radiation \citep{tavecchio00,celotti01}, or synchrotron emission of a distinct electron population \citep{stawarz02,stawarz04}. The former model requires highly relativistic bulk velocities of the jet plasma on tens- and hundreds-of-kpc scales (bulk Lorentz factors $\Gamma_j \sim 10$) or very weak jet magnetic fields, while the latter scenario assumes an efficient \emph{in-situ} acceleration process maintaining high energies of radiating particles (electron Lorentz factors $\gamma_e \geq 10^7$) along the outflows. We note that the X-ray jet emission in low-power radio galaxies is typically consistent with the extrapolation of the concave synchrotron continua from lower frequencies \citep[see][]{worrall09}.

Several crucial observational findings collected during the last decade points toward the synchrotron interpretation of the X-ray emission of large-scale quasars jet; these are summarized and discussed in more detail in section \S\,\ref{S:discussion} below. Still, the exact nature of the particle acceleration process(es) involved remains under debate. Part of the difficulty here is due to the fact that the main physical parameters of the discussed objects --- and in particular their bulk velocity profiles, content, and magnetization --- are largely unconstrained. A detailed multi-wavelength analysis of the morphological and spectral properties of large-scale quasar jets are therefore required, but as mentioned before such can be performed in reality only for a few brightest and good resolved targets.

In this paper we analyze the available best-quality radio, far-ultraviolet, and X-ray data for a particularly prominent jet in the core-dominated quasar 3C\,273 (see \S\,\ref{S:data} below). The jet has been studied in the past at various frequencies by a number of authors \citep[e.g.,][]{conway93,conway94,bahcall95,roeser91,roeser96,roeser00,neumann97,marshall01,sambruna01,jester01,jester02,jester05,jester06,jester07,martel03,uchiyama06}. The novelty of the investigation presented here lies in a careful image analysis involving combined, high-photon statistics datasets, including image deconvolution and forward-fitting of a multi-component image models (see \S\,\ref{S:analysis}). These allowed us an original insight into the multiwavelength morphology of the target, and hence into the energy dissipation processes and structure of large-scale quasar jets in general (see \S\,\ref{S:discussion}).

\begin{figure}
    \centering
    \includegraphics[width=0.45\textwidth]{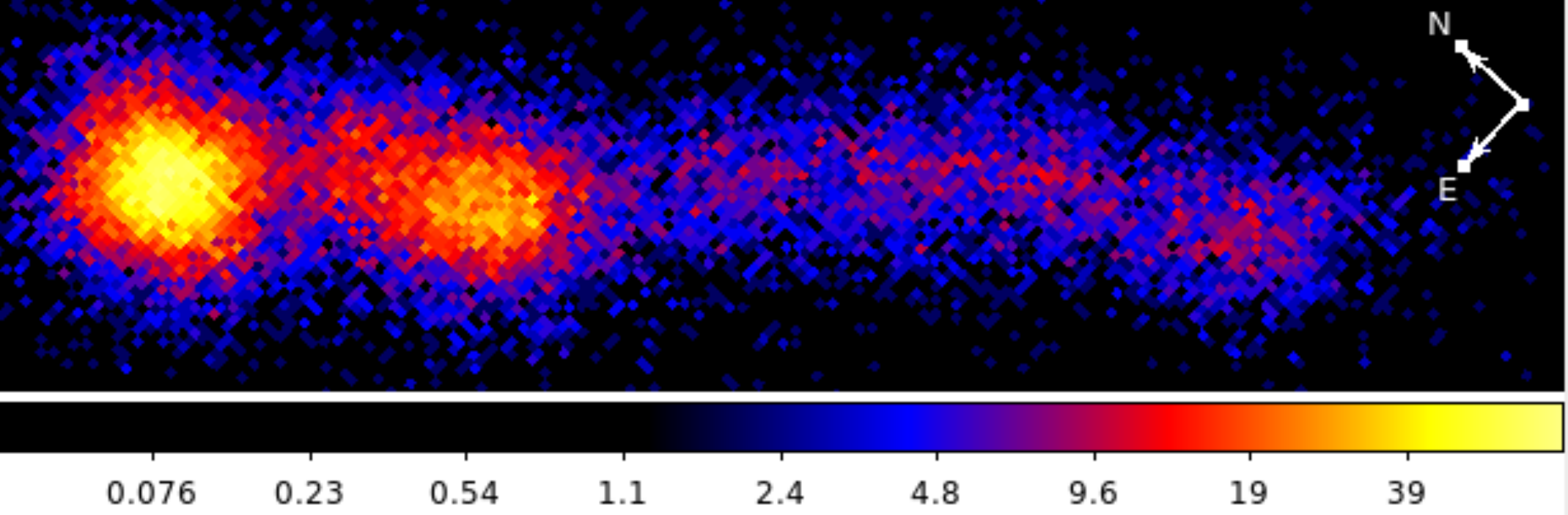}\\
    \includegraphics[width=0.45\textwidth]{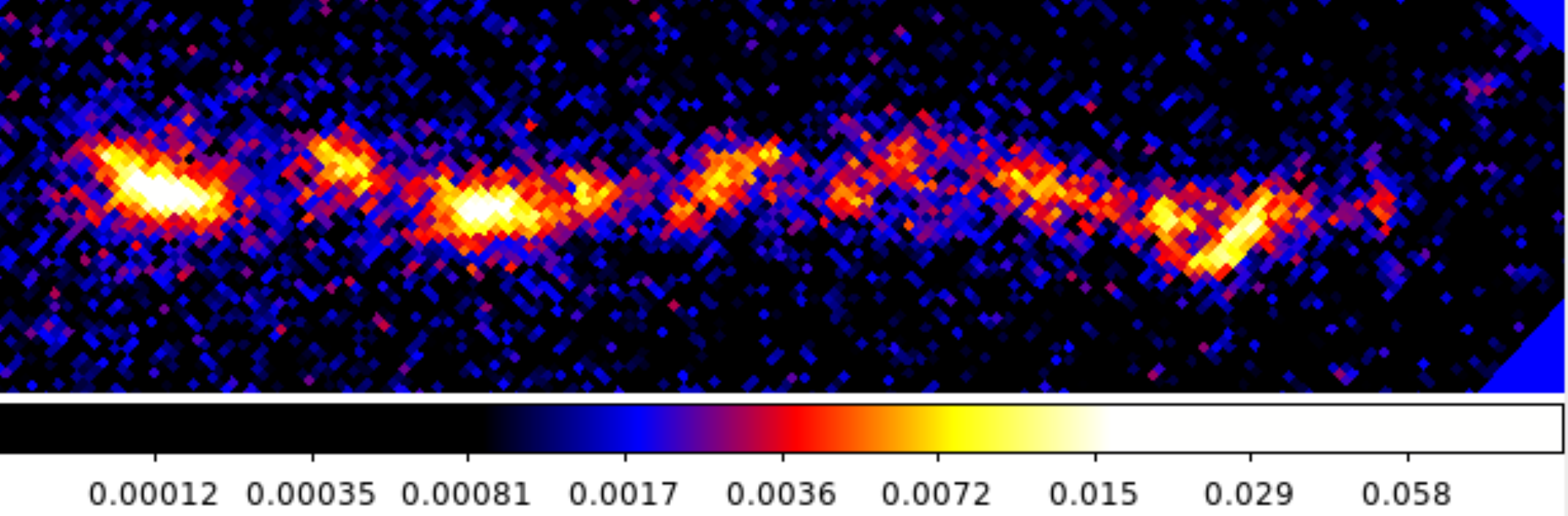}\\
    \includegraphics[width=0.45\textwidth]{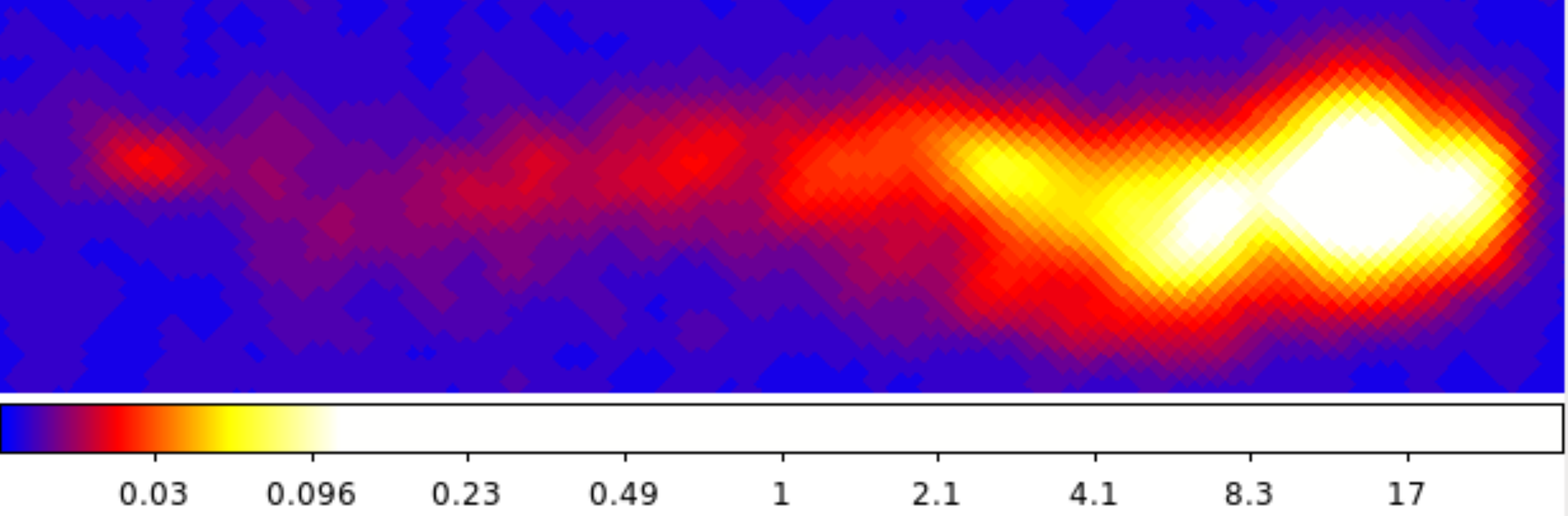}
    \caption{The {\it Chandra} ACIS X-ray, {\it Hubble} ACS/SBC ultraviolet, and {\it VLA} 8.4\,GHz radio images of the 3C\,273 jet (top, middle, and bottom panels, respectively). All maps were rebined to have the same pixel size of $0.0615\arcsec$. The units of these maps are ``counts'' (X-ray map), ``electrons/s'' (UV map), and ``Jy/beam'' (radio map).}
    \label{fig:data_all}
\end{figure}

Throughout the paper we assume a standard cosmology with $H_0=73$\,km\,s$^{-1}$\,Mpc$^{-1}$, $\Omega_{\rm m}=0.27$, and $\Omega_{\Lambda}=0.73$, so that the redshift of the target $z = 0.158$ corresponds to the luminosity distance of $d_L = 734$\,Mpc and the conversion scale of $2.647$\,kpc/$\arcsec$.

\section{Multiwavelength Data}
\label{S:data}

\subsection{{\it Chandra} X-ray Observatory}
\label{S:chandra}

We have analyzed all the available {\it Chandra} Advanced CCD Imaging Spectrometer (ACIS) data for the large-scale jet in the quasar 3C\,273. These include {\it Chandra} calibration observations as well as dedicated pointings, all listed in Table\,\ref{tab:data_xray}. The merged {\it Chandra} data binned with a factor of 0.125, which corresponds to a pixel size of $0.0615\arcsec$, are shown in the top panel of Figure\,\ref{fig:data_all}.

As the different pointings analyzed here have different off-axis angle $\theta$, and therefore various {\it Chandra} point spread function (PSF) depending on $\theta$, we have also extracted a subset of the data characterized by the best spatial resolution. In particular, we have selected 13 observations with $\theta \lesssim 0.5'$ (see Table\,\ref{tab:data_xray}). This subset was then used only for the analysis of a small-scale X-ray sub-structure of the 3C\,273 jet, in which case the size and the exact shape of the PSF are crucial; for other types of the analysis we have utilized the entire available {\it Chandra} dataset. 

\begin{table}[!t]
    {\scriptsize
    \begin{center}
        \caption{Observation log for the analyzed {\it Chandra} ACIS data.}
        \label{tab:data_xray}
        \begin{tabular}{llll}
            \hline\hline
            ObsID &  Off-axis angle $\theta$ & Start Date & Exposure \\ 
            & arcmin & UT & ksec \\
            \hline
            1198  & 0.70  & 2000-01-09 19:17:42 & 38.2 \\ 
            14455 & 0.25  & 2012-07-16 11:04:29 & 29.5 \\ 
            1711  & 1.49  & 2000-06-14 05:13:19 & 27.1 \\ 
            1712  & 0.28  & 2000-06-14 13:43:27 & 27.4 \\ 
            2463  & 0.29  & 2001-06-13 06:41:21	& 26.7 \\ 
            2464  & 1.45  & 2001-06-13 15:53:26	& 29.5 \\ 
            2471  & 1.45  & 2001-06-15 20:09:49	& 24.9 \\ 
            3456  & 0.29  & 2002-06-05 10:03:12	& 24.5 \\ 
            3457  & 0.29  & 2002-06-05 17:19:11	& 24.8 \\ 
            3573  & 0.29  & 2002-06-06 00:43:51	& 29.7 \\ 
            3574  & 1.46  & 2002-06-04 04:04:28	& 27.3 \\ 
            4430  & 0.28  & 2003-07-07 12:08:58	& 27.2 \\ 
            4431  & 1.45  & 2003-07-07 20:15:12	& 26.4 \\ 
            459   & 0.51  & 2000-01-10 06:46:11 & 38.7 \\ 
            4876  & 0.94  & 2003-11-24 23:08:40 & 37.5 \\ 
            4877  & 0.94  & 2004-02-10 03:40:39 & 34.9 \\ 
            4878  & 0.88  & 2004-04-26 20:55:16 & 34.1 \\ 
            4879  & 0.88  & 2004-07-28 03:35:33 & 35.6 \\ 
            5169  & 0.29  & 2004-06-30 12:39:18	& 29.7 \\ 
            5170  & 1.46  & 2004-06-30 21:29:53	& 28.4 \\ 
            7364  & 0.35  & 2007-01-15 08:35:19 & 2.0  \\ 
            7365  & 0.24  & 2007-07-10 21:04:28 & 2.1  \\ 
            8375  & 0.29  & 2007-06-25 05:24:03 & 29.6 \\ 
            9703  & 0.21  & 2008-05-08 21:08:12 & 29.7 \\ 
            \hline
        \end{tabular}
    \end{center}
    }
\end{table}

The analysis of the gathered {\it Chandra} data was carried out with the software package {\tt CIAO\,4.7} \citep{ciao} and the calibration database {\tt CALDB\,4.6.7}. Before the analysis the data were reprocessed using the {\tt chandra\_repro} script recommended in the {\tt CIAO} analysis threads. The pixel randomization was removed during the reprocessing. The absolute pointing accuracy of the {\it Chandra} dataset was improved in a relative sense by cross-matching one {\it Chandra} observation, ObsID 1712, chosen because of its highest number of counts ($\approx$~50\% of the total), with the other observations, prior to merging. 

\subsubsection{Instrument Response}
\label{S:response}

In the analysis of the X-ray structure of the 3C\,273 jet, which requires the best available {\it Chandra} PSF, we have used the combination of the programs {\tt ChaRT} \citep{chart} and {\tt MARX} \citep{marx} for detailed ray-trace simulation. The centroid coordinates of each selected source region were taken as a point source position for the PSF modeling. Since only the ACIS data are analyzed here, we chose the ``spectrum \& exposure time'' spectral specification in {\tt ChaRT} for the PSF modeling procedure.\footnote{\url{http://cxc.cfa.harvard.edu/chart/threads/prep/}} The background-subtracted source spectrum in the energy range $0.4 - 8.0$\,keV was created  separately for each knot and fitted using {\tt Sherpa} \citep{sherpa}, including Galactic absorption with the neutral hydrogen column density in the direction to the target $N_{\rm H} =  1.71\times10^{20}$\,cm$^{-2}$. A collection of events was made using {\tt ChaRT} by tracing rays through the {\it Chandra} X-ray optics; the rays were then projected onto the detector via {\tt MARX}, taking into account all the relevant detector issues. In this way an event file was obtained from which an image of a PSF was created. 

We note that one of the new features in the last release of {\tt MARX\,5.0.0} is the option to use the energy-dependent sub-pixel event repositioning algorithm ({\tt EDSER}) to adjust chip coordinates.\footnote{\url{http://space.mit.edu/CXC/MARX/}} However, the available version of the {\tt SAOsac ray-trace} simulator (its web interface {\tt ChaRT}, in particular) does not handle the dither motion of the telescope, and therefore currently it is not trivial to use the {\tt EDSER} algorithm in the accurate PSF modeling. Hence we have decided to use the previous version of {\tt MARX\,4.5.0} that has no {\tt EDSER} implementation.

\begin{figure}[!t]
    \centering
    \includegraphics[width=0.48\textwidth]{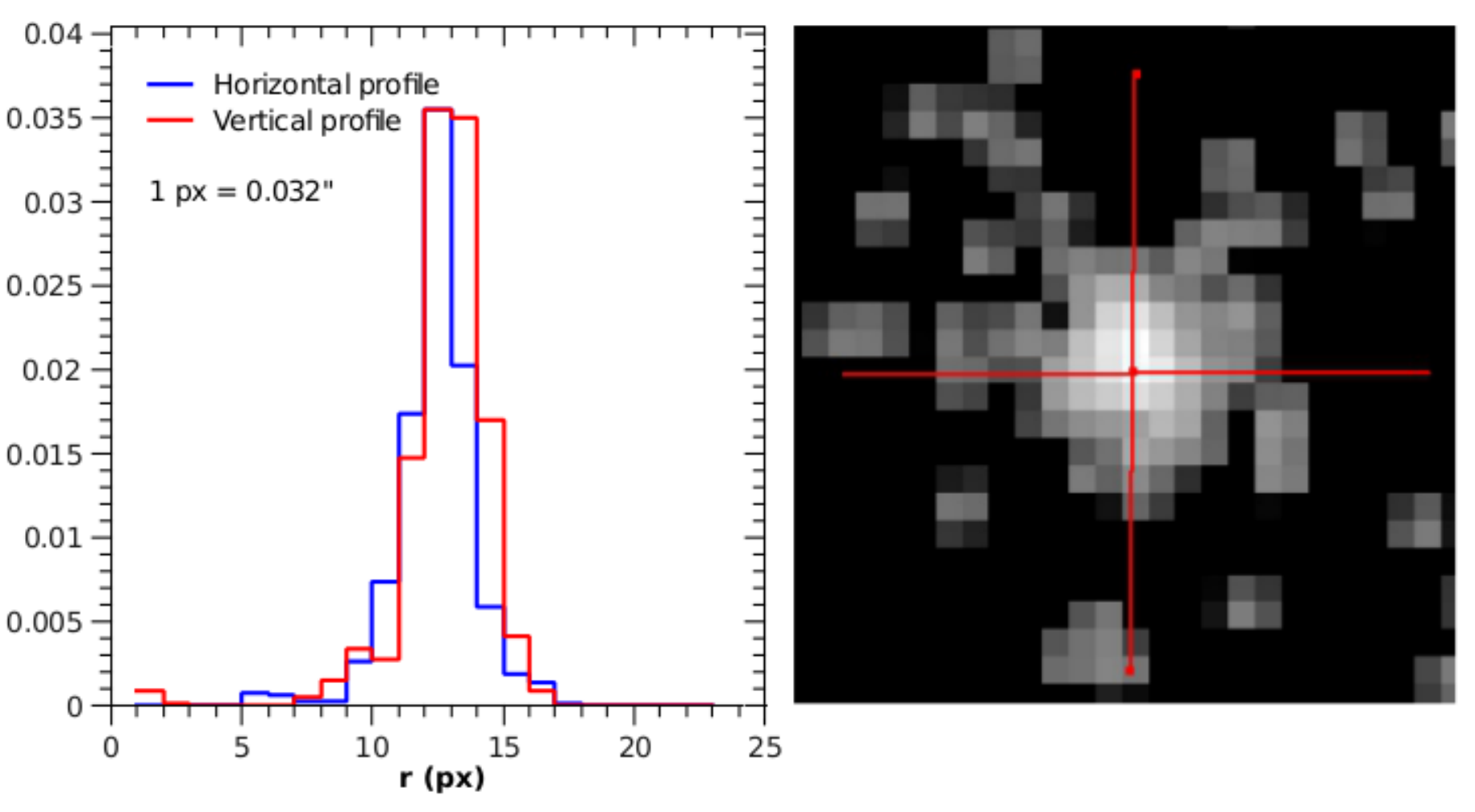}
    \caption{The FUV image of the star used for reproducing the {\it Hubble} PSF (right), along with the corresponding count profiles (left), in the original pixelization (1\,px\,$ = 0.032\arcsec$).}
    \label{fig:uv_star}
\end{figure}

Because the {\it Chandra} instrumental response depends on a number of factors (including a source position, energy range considered, an aspect solution, etc.), one has to take into account possibly uneven exposure in the map analysis. We have therefore generated the exposure map using the {\tt CIAO} script {\tt fluximage} in the range $0.4-8$\,keV with the effective energy of 2\,keV, consistent with the energy range of the analyzed dataset. The binning used for creating the exposure map was the same as for the image data (binning factor 0.125, corresponding to a pixel size of $0.0615\arcsec$). In this way we have confirmed that the exposure does not vary significantly across the entire field of interest, and that the source spatial characteristics do not change noticeably with or without the exposure map correction.

\subsubsection{Noise Estimation}
\label{S:noise}

In the analysis of the {\it Chandra} maps we take into account statistical noise only. We assume that counts are sampled from the Poisson distribution with a mean value equal to the total number of the observed counts $N_i$ in each bin. Hence the count standard deviation for a bin $i$ can be taken as $\sqrt{N_i}$. However, when the average number of counts in a given bin is small ($N_i < 5$), one cannot simply assume that the Poisson distribution --- from which the counts are sampled --- has a nearly Gaussian shape; the standard deviation for such a `low-count case' has been derived by \citet{gehrels86} as, instead,
\begin{equation}
\sigma_i = 1 + \sqrt{N_i + 0.75} \, .
\label{eq:sigma_gehrels}
\end{equation}
In the above, the higher-order terms have been omitted, so the expression is accurate to approximately $1\%$. 

Equation\,\ref{eq:sigma_gehrels} can be applied when no background subtraction is performed. Otherwise, one should use the standard error propagation $\sigma^2_{i,\,{\rm NB}} = \sigma^2_i + \sigma^2_{i,\,{\rm B}}$, where $\sigma_i$ is derived from equation\,\ref{eq:sigma_gehrels}, and $\sigma_{i,\,{\rm B}}$ is the standard deviation of the expected background at the source position (typically estimated from a region selected in the source vicinity). 

\begin{figure}[!t]
    \centering
    \includegraphics[width=0.3\textwidth]{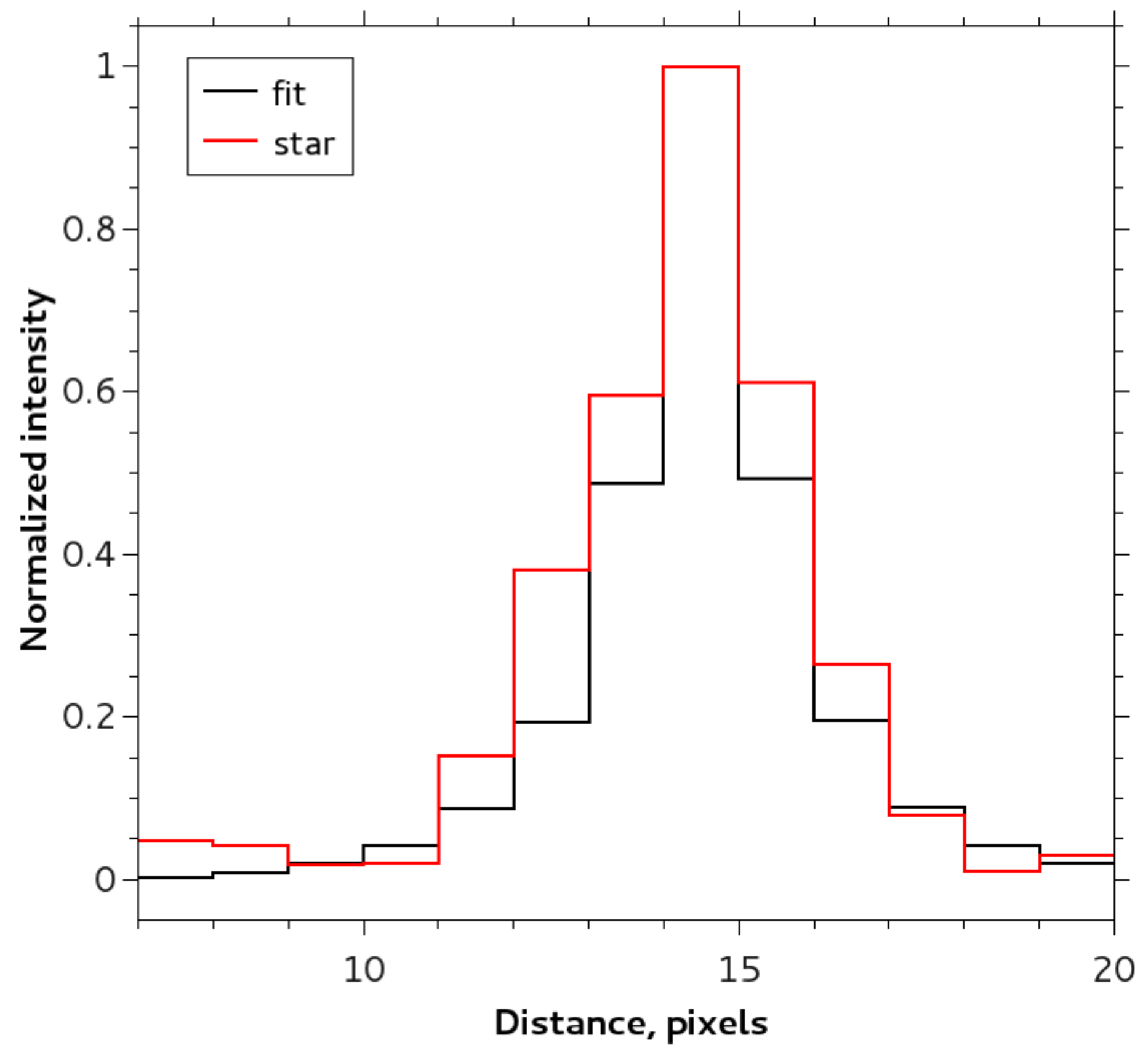}
    \caption{The FUV profile of the field star and the PSF profile obtained from the encircled energy curve.}
    \label{fig:sbc_psf}
\end{figure}

\subsection{Hubble Space Telescope}
\label{S:hubble}

In addition to the {\it Chandra} observations described above, we have also analyzed the far-ultraviolet (FUV) data for 3C\,273, obtained using the Advanced Camera for Surveys (ACS)/Solar Blind Channel (SBC) onboard the {\it Hubble} Space Telescope at $\simeq 150$\,nm \citep[][see Table\,\ref{tab:data_uv}]{jester07}. The resulting FUV image of the jet with the $0.0615\arcsec$ binning is presented in the middle panel of Figure\,\ref{fig:data_all}.
 
\begin{table}[!b]
    {\scriptsize
    \begin{center}
        \caption{Observation log for the analyzed {\it Hubble} ACS/SBC data in the \texttt{F150LP} filter.}
        \label{tab:data_uv}
        \begin{tabular}{llll}
            \hline\hline
            Target & Dataset & Date & Exposure \\ 
            & & UT & ksec\\
            \hline
            J122903+020318 & J8P001010 & 04/08/2004 & 0.9\\
            3C273-JET & J8P001TSQ & 04/08/2004 & 1.6\\
            3C273-JET & J8P001020 & 05/08/2004 & 2.8\\
            3C273-JET & J8P001030 & 05/08/2004 & 2.8\\
            \hline
           \end{tabular}
       \end{center}
       }
   \end{table}

\subsubsection{Point Spread Function}
\label{S:psf}
 
Before utilizing the {\it Hubble} data in the analysis of the small-scale structure of the target, the telescope blurring has to be reduced, and this can be done by means of a deconvolution. This procedure requires a good characterization of the corresponding PSF, which can in principle be accomplished with the {\tt Tiny Tim} modeling tool. However, here we cannot use this tool as it does not handle the multidrizzled ACS/SBC files. Instead, we have estimated the PSF using the image of a point-like star in the dataset J8P001010 (see Table\,\ref{tab:data_uv} and Figure\,\ref{fig:uv_star}); the same star has been used for the astrometry calibration by \citet{jester07}. 

In order to investigate further if the selected star reproduces adequately the PSF of the instrument, we have performed the additional comparison analysis using the information on the encircled energy fraction for the ACS/SBC, generated from the data acquired during the servicing mission orbital verification.\footnote{see the {\it Hubble} ACS Handbook for Cycle\,12 at\\ \url{http://documents.stsci.edu/hst/acs/documents/handbooks/cycle12/}} The radial distribution of the encircled energy fraction is 
\begin{equation}
\zeta(r) = \frac{\int_0^r \, f\!(r') \, 2\pi r' dr'}{\int_0^{r_{\rm max}} \, f\!(r') \, 2\pi r' dr'} \, ,
\label{eq:nu_f}
\end{equation}
for a given PSF profile $f(r)$. The above relation allows one to restore the PSD profile from the fitted $\zeta(r)$ curve.

Our analysis indicates that the PSF profile can be approximated by a Gaussian function only in the innermost region $\leq 0.05\arcsec$, and that the wings of the PSF clearly deviate from a Gaussian shape. A better agreement is obtained assuming a Lorentz function profile for the PSF, although significant deviations are still present. Acceptable fits are obtained instead by adding a polynomial of the 6th order to the Lorentz function. We have therefore adopted this parametrization for creating the FITS-file of the PSF; with such, the FUV image of the star is consistent with the PSF, as shown in Figure\,\ref{fig:sbc_psf}.

The estimated FWHM of PSF obtained from image of star and encircled energy fraction are $0.13\arcsec$ and $0.18\arcsec$ respectively (small difference caused different binning).

\subsection{Very Large Array}
\label{S:vla}

The {\it Chandra} X-ray and {\it Hubble} FUV data for the 3C\,273 jet have been augmented by the archival NRAO\footnote{The National Radio Astronomy Observatory is a facility of the National Science Foundation operated under cooperative agreement by Associated Universities, Inc.} Very Large Array ({\it VLA}) observations performed in 1995 at 8.4\,GHz (project code AR\,334). This {\it VLA} map was kindly provided by R. A. Perley and has a clean beam
of FWHM = 0.35$\arcsec$.
The resulting radio image of the jet is shown in the bottom panel of Figure\,\ref{fig:data_all}. Its rms level is 0.44 mJy/beam and the dynamic range is 1:75000.

\section{Data analysis}
\label{S:analysis}

\subsection{X-ray Structure of the Jet}
\label{S:xray_structure}

\subsubsection{Forward-fitting of Multi-component Source}
\label{S:forward}

Mapping of an extended source characterized by a spatial photon distribution $s_i(x, y)$ with a telescope of a given PSF $P_i(x, y)$ results in a source image $N_i(x, y)$ that can be represented as a convolution 
\begin{eqnarray}
N_i(x,y) & = & \sum_{x'}\sum_{y'} \,\, s_i(x',y') \,\, P_i(x-x',y-y') \nonumber \\ 
& \equiv & s_i \circ P_i \, ,
\label{eq:conv}
\end{eqnarray}
where the index $i$ corresponds to the bin with coordinates $(x,y)$. If the exact source photon distribution $s_i(x,y)$ is unknown, one can assume a source model $S_i$, and build the model of a source image as
\begin{equation}
N^{\star}_i = S_i \circ P_i \, .
\label{eq:jet_model_general}
\end{equation}
In a general case, the source model $S_i$ can be written as a sum of $n$ components $G_{i, k}$ plus a uniform background $C$, namely
\begin{equation}
S_i = \sum_{k=1}^n G_{i, k} + C \, .
\label{eq:jet_source}
\end{equation}
With such, the source image model (\ref{eq:jet_model_general}) becomes
\begin{equation}
N^{\star}_i = \sum_{k=1}^n \left(G_{i, k} \circ P_i \right) + C \, ,
\label{eq:jet_model}
\end{equation}
where $P_i$ is the PSF at the position of a component $G_{i, k}$. As the instrumental PSF may depend on the spatial coordinates and photon energy, the PSF modeling has to be performed for each source component separately.

\begin{table}
    {\scriptsize
    \begin{center}
        \caption{Best-fit parameters for the 3C\,273 X-ray jet source model {\tt 4e3s}.} 
        \label{tab:4e3s}
        \begin{tabular}{lllll}
            \hline\hline
            Component & Major-axis & Minor-axis  & Ellipticity       & PA$^{\star}$ \\
         (knot name)  & FWHM       & FWHM        &  $\varepsilon$    & deg         \\
                      & $\mathrm{wa}$ [arcsec] & $\mathrm{wb}$ [arcsec] &  &   \\

            \hline
            G1 (A)  & $0.24 \pm 0.04$ & $0.15 \pm 0.06$ & $0.39 \pm 0.11$ &  $11.43 \pm 7.62$  \\   
            G2  	& $0.17 \pm 0.03$ & $0.17 \pm 0.03$ &  0.0            &  --               \\
            G3 (B2) & $0.45 \pm 0.05$ & $0.11 \pm 0.02$ & $0.75 \pm 0.02$ & $63.07 \pm 1.53$   \\ 
            G4 (C1) & $1.05 \pm 0.16$ & $0.29 \pm 0.14$ & $0.73 \pm 0.10$ & $53.45 \pm 7.84$   \\ 
            G5	    & $0.50 \pm 0.10$ & $0.50 \pm 0.10$ &  0.0            &  --               \\
            G6      & $0.13 \pm 0.05$ & $0.13 \pm 0.05$ &  0.0            &  --               \\
            G7 (H3) & $1.18 \pm 0.08$ & $0.11 \pm 0.07$ & $0.9 \pm 0.06$  & $44.85 \pm 3.20$   \\
            \hline
        \end{tabular}
    \end{center}
    $^{\star}$ The jet PA $\approx 40$ deg;
    }
\end{table}

The residuals between the analyzed image model (\ref{eq:jet_model}) and the data, i.e. the residuals between the number of source counts $N_i$ in a given image bin $i$ and the number of modeled source counts $N^{\star}_i$ in the same bin, can next be investigated in terms of a standard deviation of the image count distribution given in equation\,\ref{eq:sigma_gehrels}. This results in a $\sigma$-map calculated as
\begin{equation}
R_i = \frac{N_i - N^{\star}_i}{\sigma_i} \, ,
\label{eq:sigmap}
\end{equation}
which reveals the regions with statistically significant disagreement between the data and the model.

\begin{figure*}[!t]
    \centering
    \includegraphics[width=0.33\textwidth]{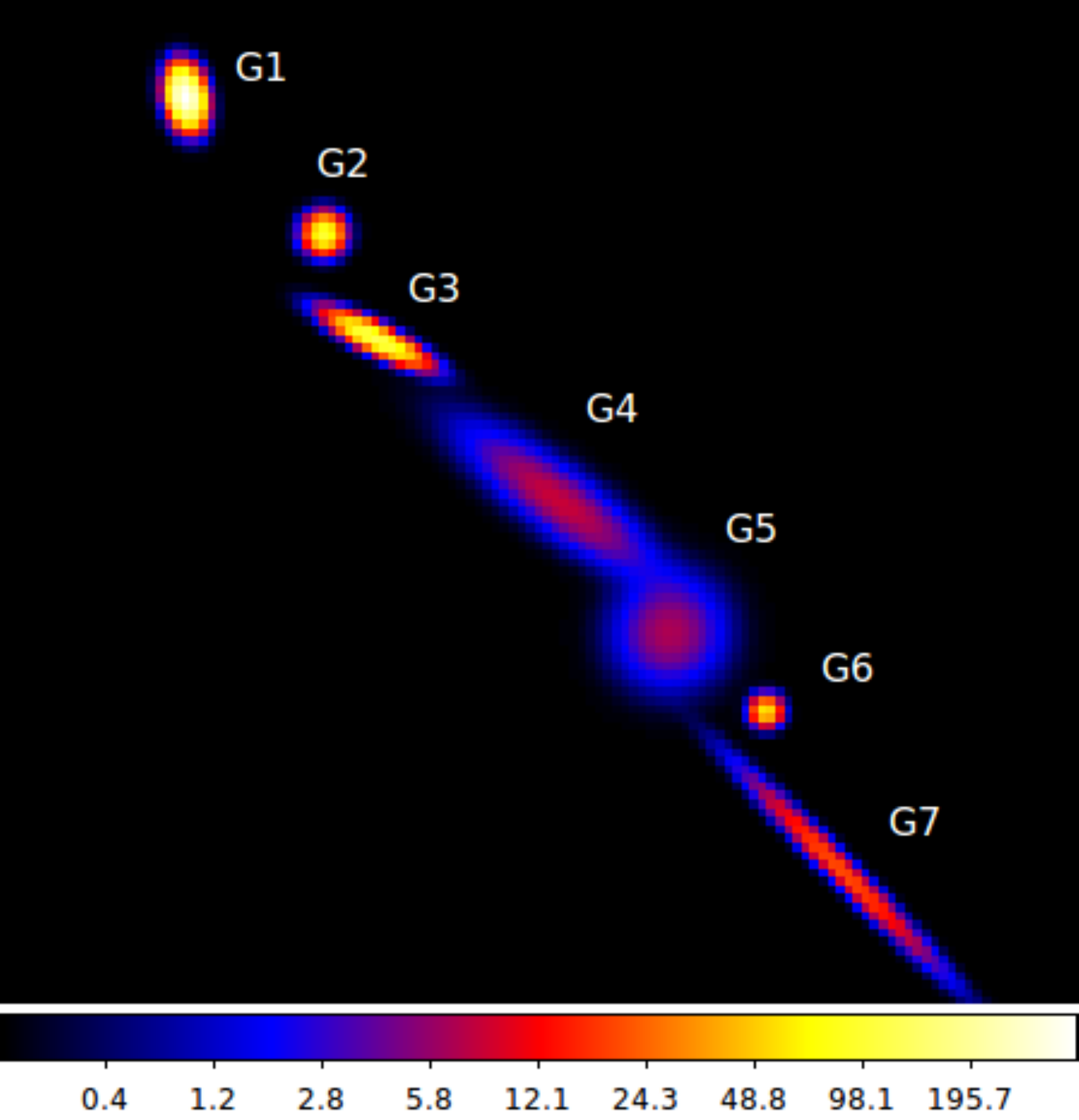}
    \includegraphics[width=0.33\textwidth]{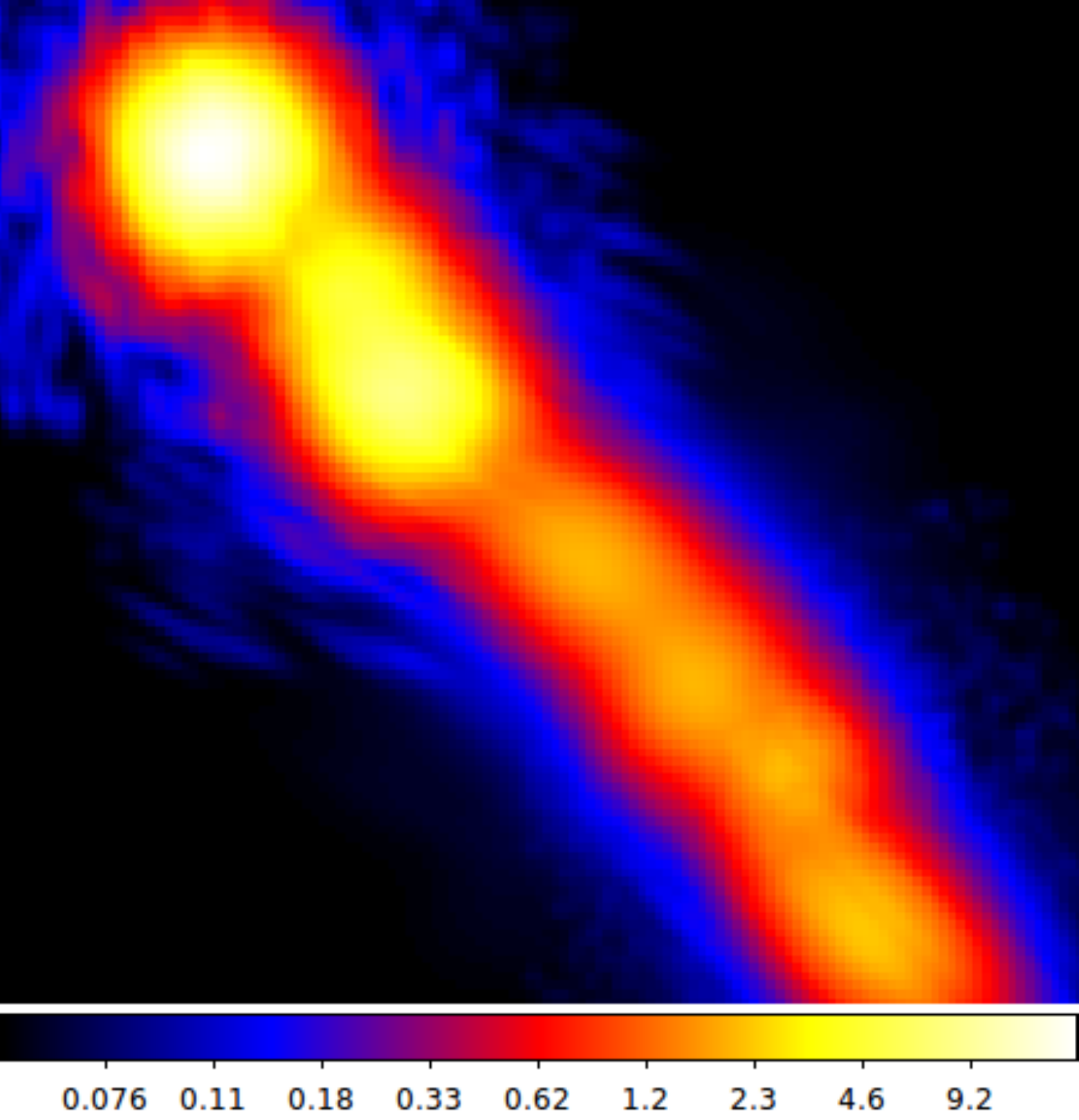}
    \includegraphics[width=0.33\textwidth]{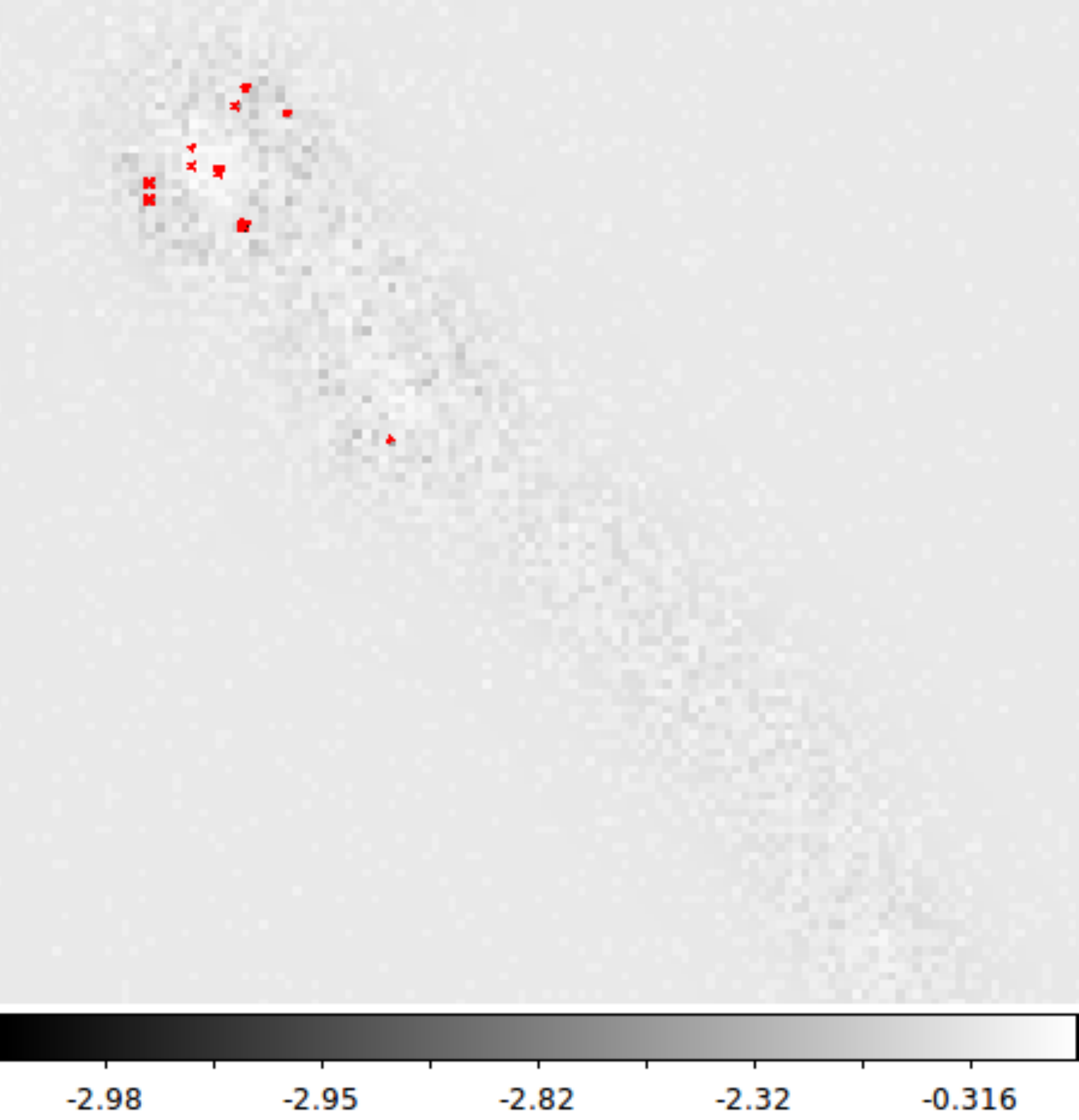}
    \caption{Source image fitting for the X-ray jet in 3C\,273: the source model {\tt 4e3s} (left), the convolved jet model (middle), and the $\sigma$-map (right). In the latter panel, red contours correspond to $2\,\sigma$ residuals.} 
    \label{fig:4e3s}
\end{figure*}

In order to find the best fit parameters for the source model $S_i$ in the particular case of the 3C\,273 jet analyzed here, we applied the forward-fitting algorithm from the {\tt Sherpa} software package; the maximum likelihood estimation (MLE) method was applied using the Nelder-Mead Simplex optimization algorithm based on Cash maximum likelihood function $C = 2\sum[M_i-D_i\ln M_i]$, where $M_i$ and $D_i$ are, respectively, a model amplitude and a number of observed counts in a given bin $i$.  The initial number and positions of possible source components $G_{i, k}$ for the fitting were estimated by adaptively smoothing the jet image using the {\tt CIAO} tool {\tt csmooth}, with the minimal significance of the signal under the kernel set at the value of 3. This allowed us to identify statistically significant brightness enhancements, which were next visually confirmed as distinct source components. 

Taking into account that, in the case of a low number of counts, particular realisations of the PSF image may vary between different simulations due to photon fluctuations, in our analysis we used only the PSF image obtained with a maximum number of counts allowed by the ray-trace simulator --- i.e. the best possible PSF image --- for a given observation. The influence of the related uncertainties in the PSF shape on the source forward-fitting is analyzed in Appendix~\ref{sec:app1}: as shown, standard deviations cased by fluctuations in the PSF shape are comparable with the errors of the forward-fitting procedure.

We have investigated several source models for the X-ray jet in 3C\,273, starting from the simplest model with all the knots assumed to be spherical, and including also more complicated scenarios with elliptical knots as determined using the Akaike and Bayesian information criteria (see Appendix~\ref{sec:app2} for a detailed discussion). All in all, we have selected the best model {\tt 4e3s}, consisting of four elliptical (G1, G3, G4 and G7) and three circular (G2, G5 and G6) Gaussian components, plus a constant background (C). All the model parameters have been adjusted during the fitting procedure. The results are summarized in Table\,\ref{tab:4e3s}, and visualized in Figure\,\ref{fig:4e3s}.

In the source image fitting procedure one should, in principle, include also an exposure map to account for an uneven instrumental cover. On the other hand, as noted in \S\,\ref{S:response}, the exposure map for the {\it Chandra} data analyzed here is almost constant, with only small variations not exceeding $1\,\%$. Hence we conclude that adding the exposure map in the image fitting would not affect significantly the resulting spatial distribution of source counts. This simplification is however justified only when the spatial structure of the source is considered, but not when the physical flux analysis is being performed.

One should finally note that the forward-fitting procedure works well for the X-ray--bright segment of the 3C\,273 jet, but does not reproduce adequately the component sizes in the cases of the outer, fainter parts, due to the limited photon statistics. Therefore, another technique has to be applied to restore the X-ray image of the entire jet. Still, the multi-component forward-fitting analysis of the {\it Chandra} data presented in this section indicates robustly that at least the most prominent brightness enhancements in the X-ray jet of 3C\,273 --- the ``knots'' --- are not point-like, and can be resolved as extended features with sizes of about $0.1\arcsec - 0.3\arcsec$.

\subsubsection{Lucy-Richardson Deconvolution Algorithm}
\label{S:deconvolution}

As mentioned in the previous section, an image of an astronomical source can be described mathematically as a convolution of the intrinsic source brightness distribution with the instrument blurring function, the PSF. Therefore, one of the possibilities for restoring the intrinsic source distribution from the observed image is to use one of the established deconvolution techniques. In the particular case of the 3C\,273 jet studied here, we have applied the Lucy-Richardson Deconvolution Algorithm (LRDA), which is implemented in the {\tt CIAO\,4.6} tool {\tt arestore}.\footnote{\url{http://cxc.harvard.edu/ciao/ahelp/arestore.html}} This algorithm requires an image of a PSF, and so in our analysis we used the modeled {\it Chandra} PSF generated by the {\tt ChaRT} and {\tt MARX} programs for detailed ray-trace simulations \citep[][see \S\,\ref{S:response}]{chart,marx}. 

\begin{figure}[!t]
    \centering
    \includegraphics[width=0.45\textwidth]{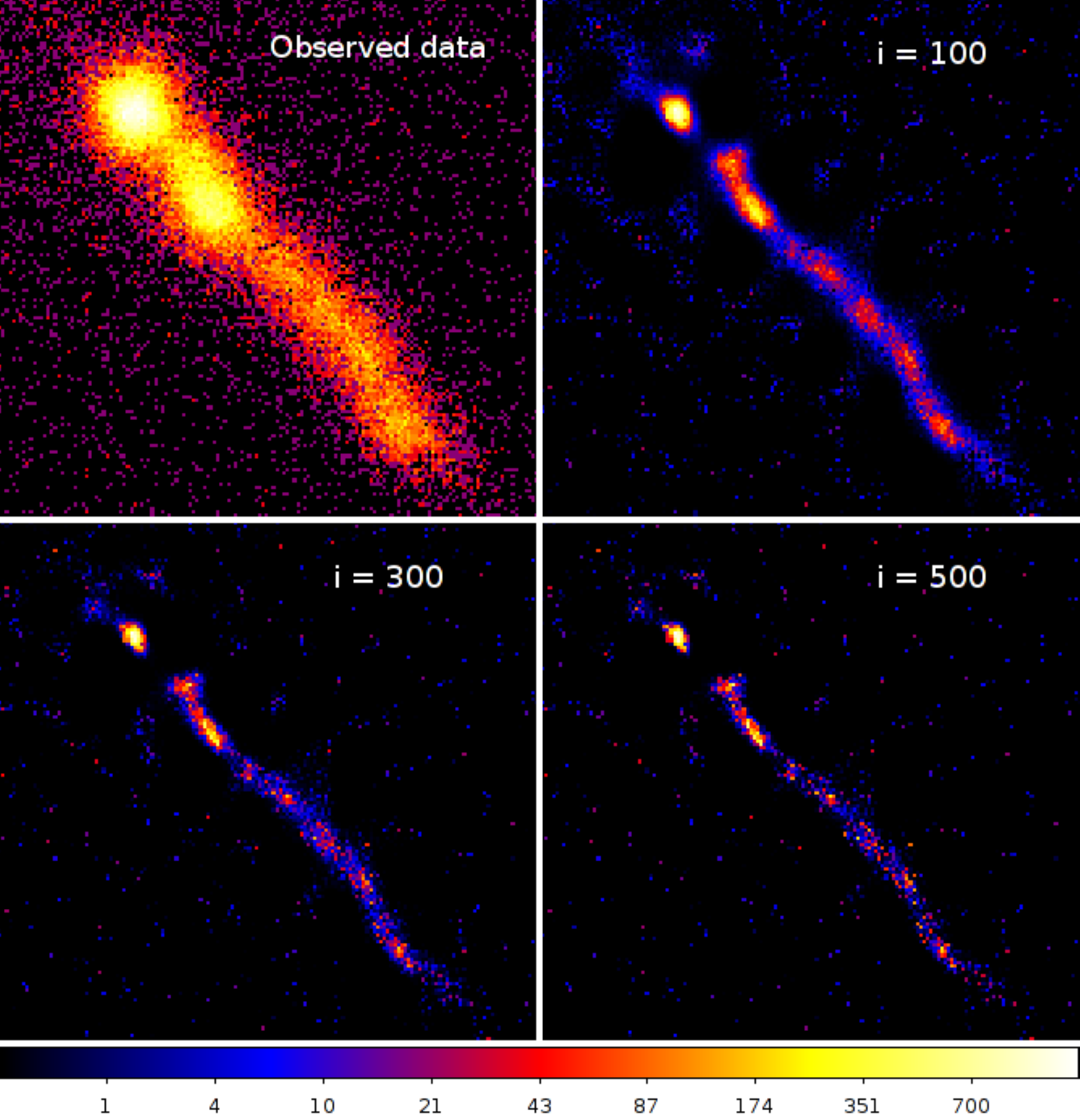}
    \caption{The observed image of the X-ray jet in 3C\,273, and the exemplary restored images for different numbers of iterations in the LRDA.}
    \label{fig:restored_examples}
\end{figure}

It should be emphasized that the applied method may not return reliable convergence and uncertainty information, and users should be rather cautious in interpreting the results of the deconvolution, especially when dealing with faint and/or extended sources. Also, the characteristics of the restored image (component sizes, amplitudes, etc.) may vary for different numbers of iterations performed. This effect is visualized in Figure\,\ref{fig:restored_examples}, which shows the resulting deconvolved X-ray image of the 3C\,273 jet for three exemplary numbers of iterations set as input parameters in the LRDA.

In order to choose a reasonable number of iterations assuring that the results of the performed deconvolution are valid, we have performed the following comparison analysis between the restored images and the data. First, we constructed the model as a convolution
\begin{equation}
M_i = S_i \circ P_i \, ,
\end{equation}
where $M_i$ and $S_i$ are the numbers of counts per bin $i$ in the model and in the restored image, respectively, while $P_i$ is the corresponding PSF. Next, we evaluated the reduced $\chi^2$ statistic for different numbers of iterations,
\begin{equation}
\chi^2/ \mathrm{dof} = (1/\mathrm{dof}) \,\, \sum_i \frac{(N_i - M_i)^2}{\sigma_i^2} \, ,
\end{equation}
where $N_i$  is the total number of the observed counts per bin $i$, $\sigma_i$ is the corresponding error (for which we account for statistical noise only;  see \S\,\ref{S:noise}), and ``dof'' (degrees of freedom) is the number of bins in a given region. 

\begin{figure}[!t]
    \centering
    \includegraphics[width=0.48\textwidth]{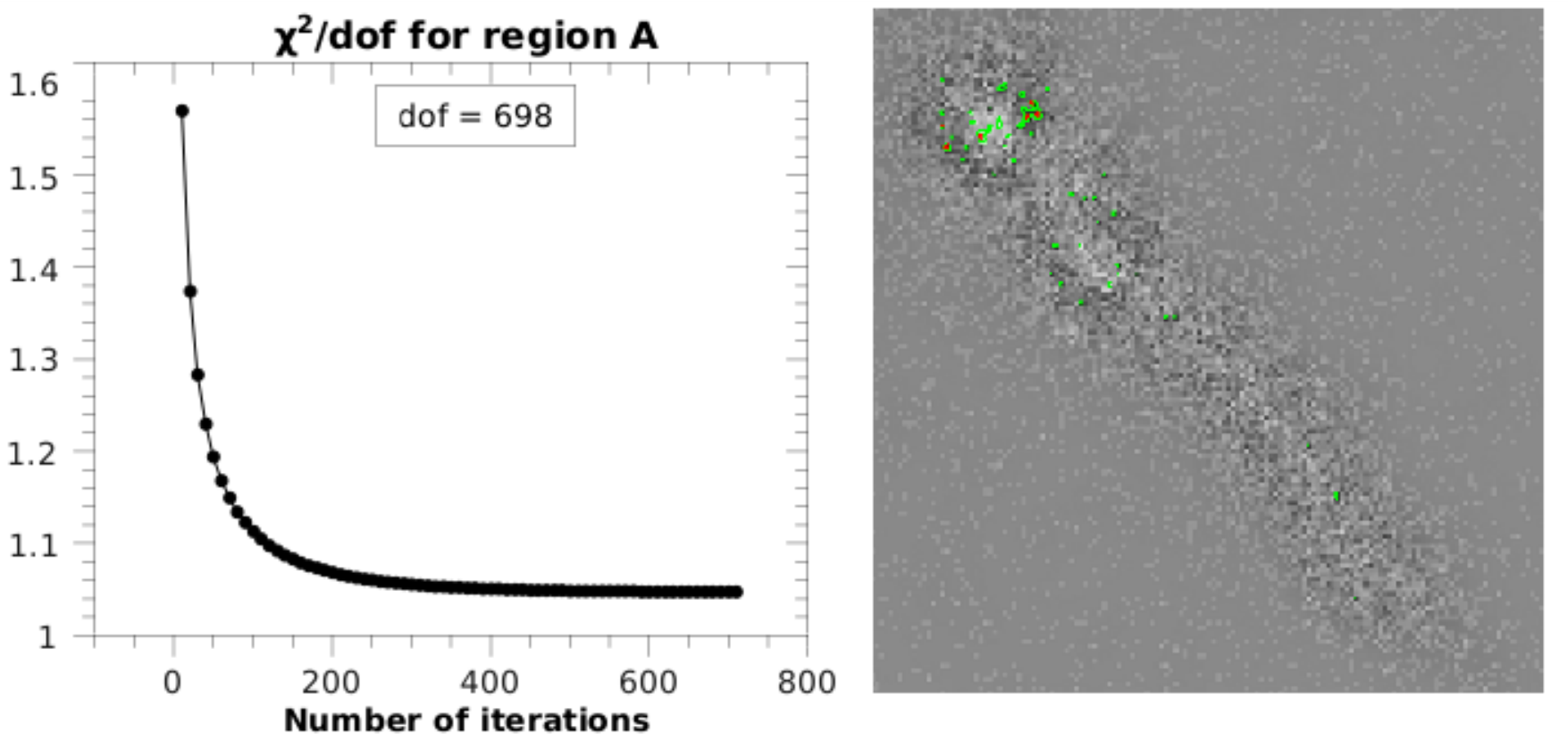}
    \caption{{\it Left:} Reduced $\chi^2$ as a function of the number of iterations in the LRDA for the ``knot A'' region. {\it Right:} Residual map for the entire jet with the optimal number of iterations (corresponding to $\chi^2/\mathrm{dof} \approx 1.1$) in the LRDA. The map is normalized by standard deviation; the green contours correspond to the $2\sigma$ deviation, and red contours to the $3\sigma$ deviation.}
    \label{fig:chi2}
\end{figure}

The above procedure has been applied to the first, X-ray-bright knot A in the 3C\,273 jet \citep[following the standard labeling adopted in the literature, see][]{jester05}, equivalent to the G1 source component from \S\,\ref{S:forward}. We have calculated the reduced $\chi^2$ for different number of iterations, as presented in the left panel of Figure\,\ref{fig:chi2}. From this, we concluded that the number of iterations should be limited to a value corresponding to $\chi^2/\mathrm{dof}=1.1$ (i.e. $\simeq 100$ in the considered case). We have then inspected the normalized distribution of residuals between the data $N_i$ and the model $M_i$, namely $R_i = (N_i - M_i)/\sigma_i$, across the entire jet image; the resulting map is shown in the right panel of Figure\,\ref{fig:chi2}.

The outcome of the {\it Chandra} data analysis performed with the LRDA as described above is consistent with the main finding following from the multi-component forward-fitting analysis discussed in \S\,\ref{S:forward}: X-ray knots in the 3C\,273 jet are not point-like, and can be resolved as extended features. Moreover, the deconvolution procedure using the LRDA, which we have also applied to the FUV data obtained with the {\it Hubble} ACS/SBC, allows us to investigate in detail the multiwavelength transverse structure of the 3C\,273 jet. The results are discussed in the following sections.

\subsection{Transverse Jet Profiles}
\label{S:transverse}

In order to investigate the transverse profiles of the 3C\,273 jet, we have utilized the original {\it VLA}, deconvolved {\it Hubble}, and {\it Chandra} maps as discussed in the previous sections. The corresponding radio and X-ray images of the outflow are presented in Figure\,\ref{fig:XR_knots_con}. Since a \emph{multiwavelength} comparison is our main objective here, in this section we analyze only the brightest parts of the jet which are clearly detected at radio, FUV, and X-ray frequencies. Using the standard 3C\,273 labeling, these brightest segments of the outflow correspond to the knots A, B2, C1, and H3 \citep[see][and Figure\,\ref{fig:XR_knots_con}]{jester05}. 

Taking into account that some of these knots do not appear symmetric, but instead display elongated, elliptical shapes, which are moreover misaligned with respect to the main jet axis, the transverse profiles were calculated through the minor axis of each knot. 

\begin{figure*}[!t]
    \centering
    \includegraphics[width=0.33\textwidth]{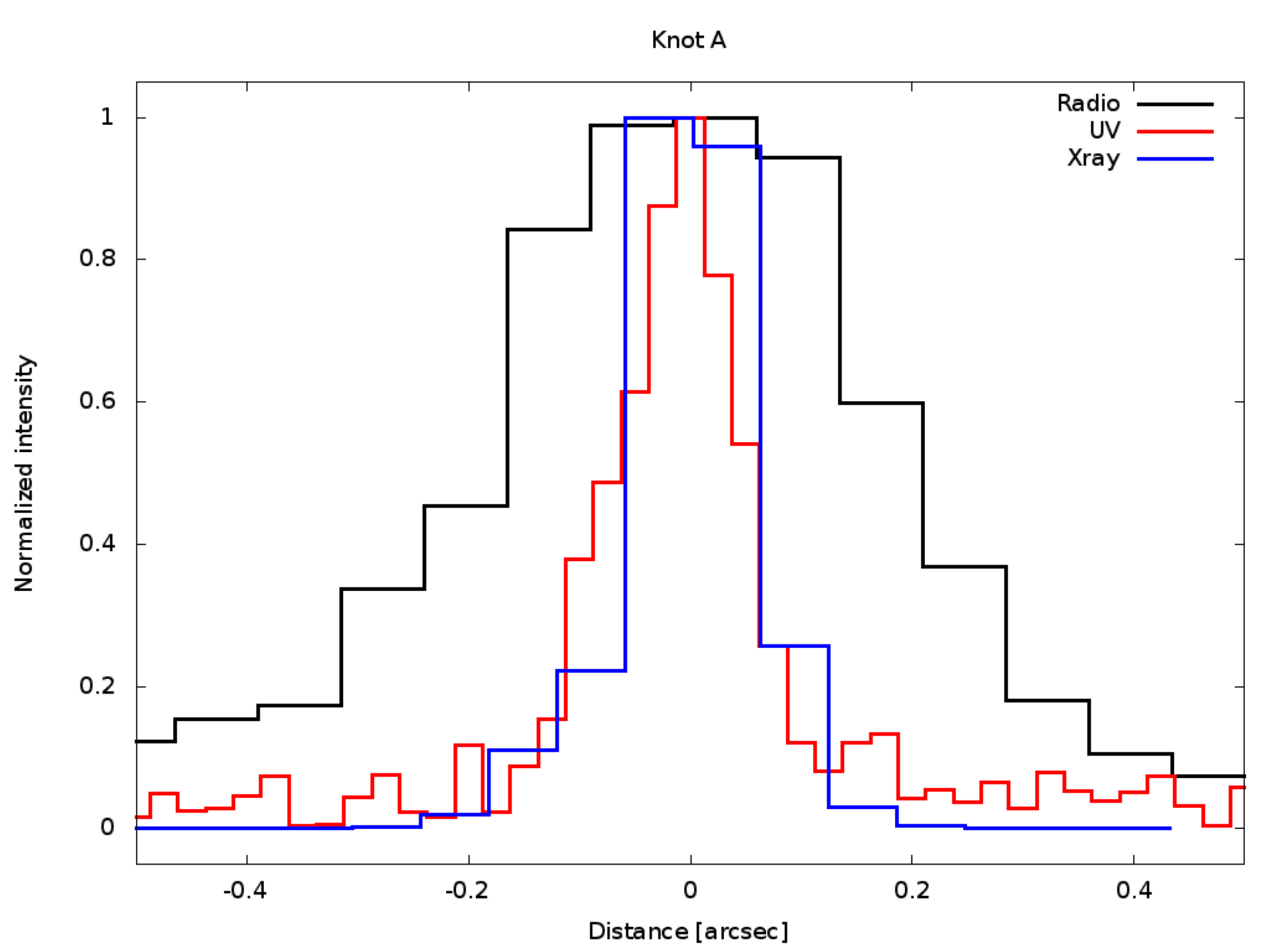}
    \includegraphics[width=0.33\textwidth]{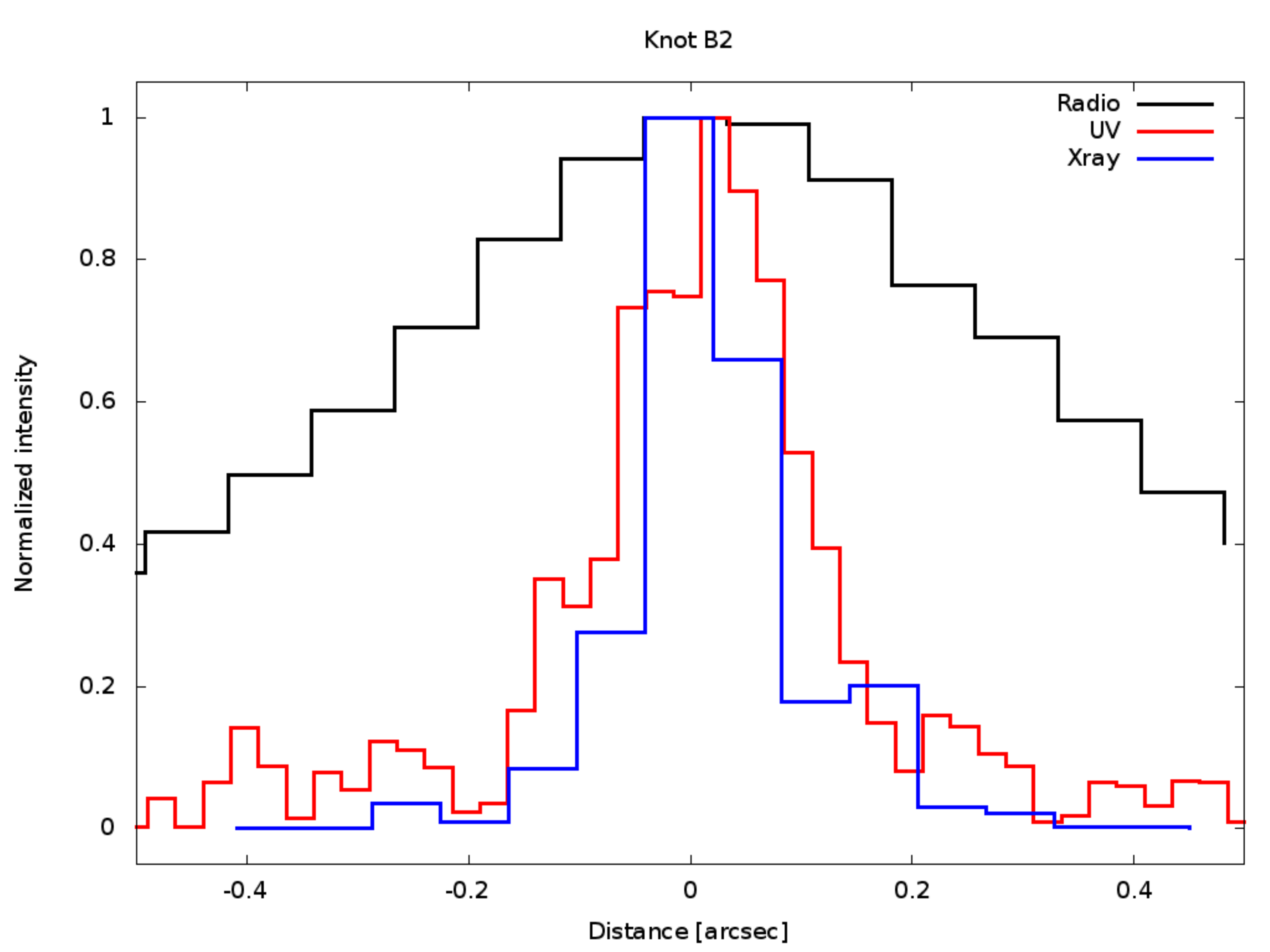}
    \includegraphics[width=0.33\textwidth]{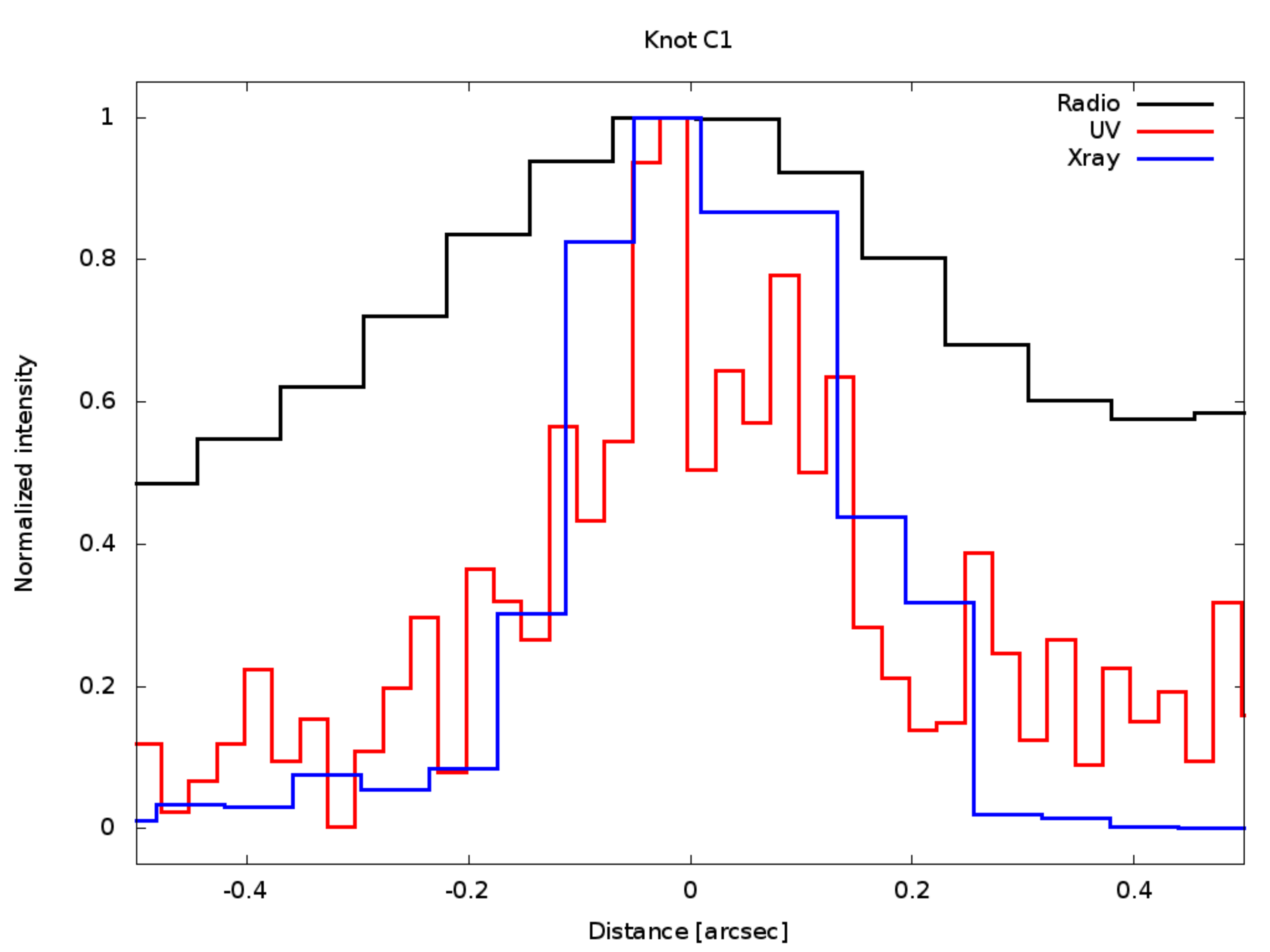}
    \includegraphics[width=0.33\textwidth]{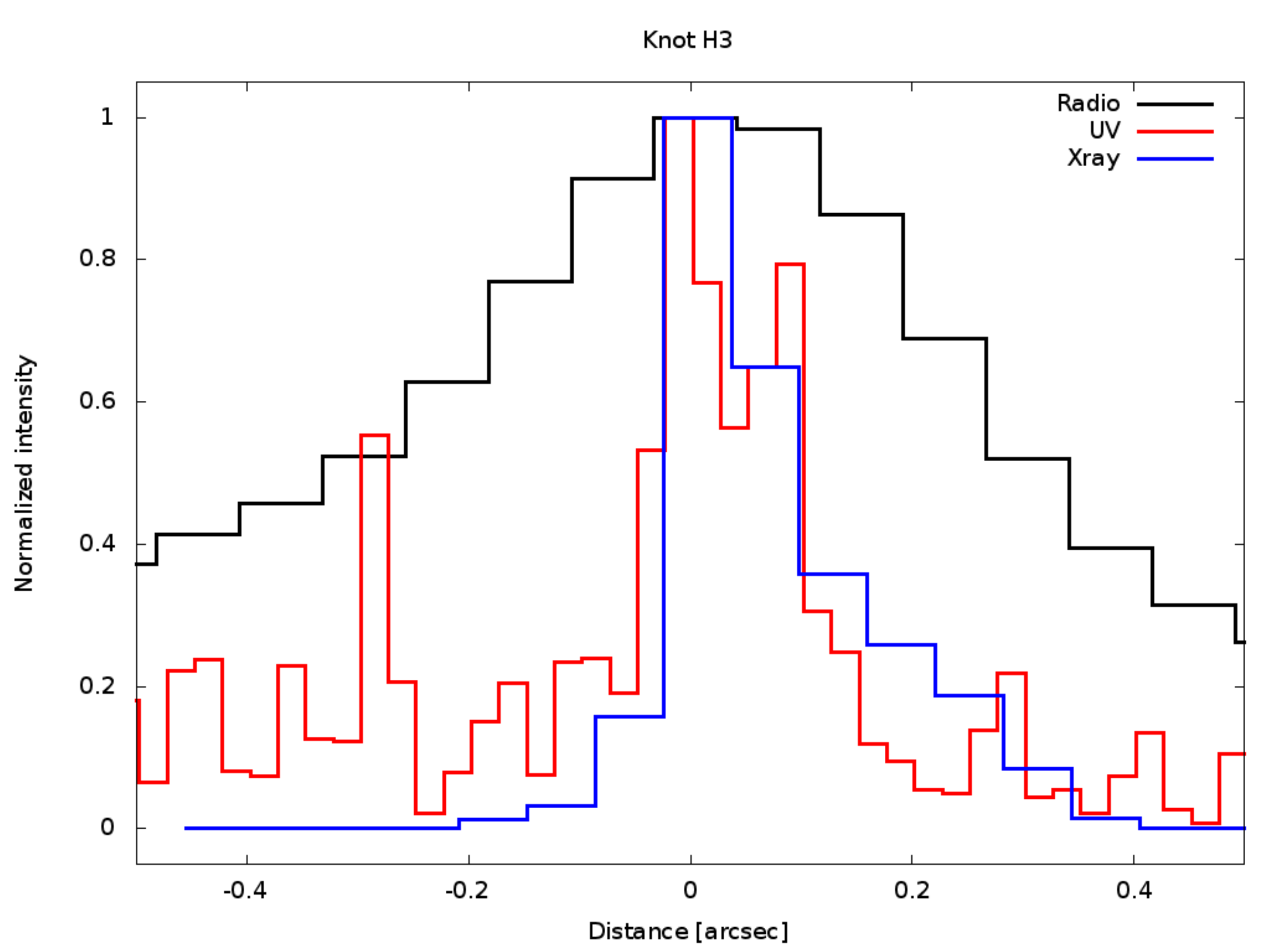}
    \includegraphics[width=0.33\textwidth]{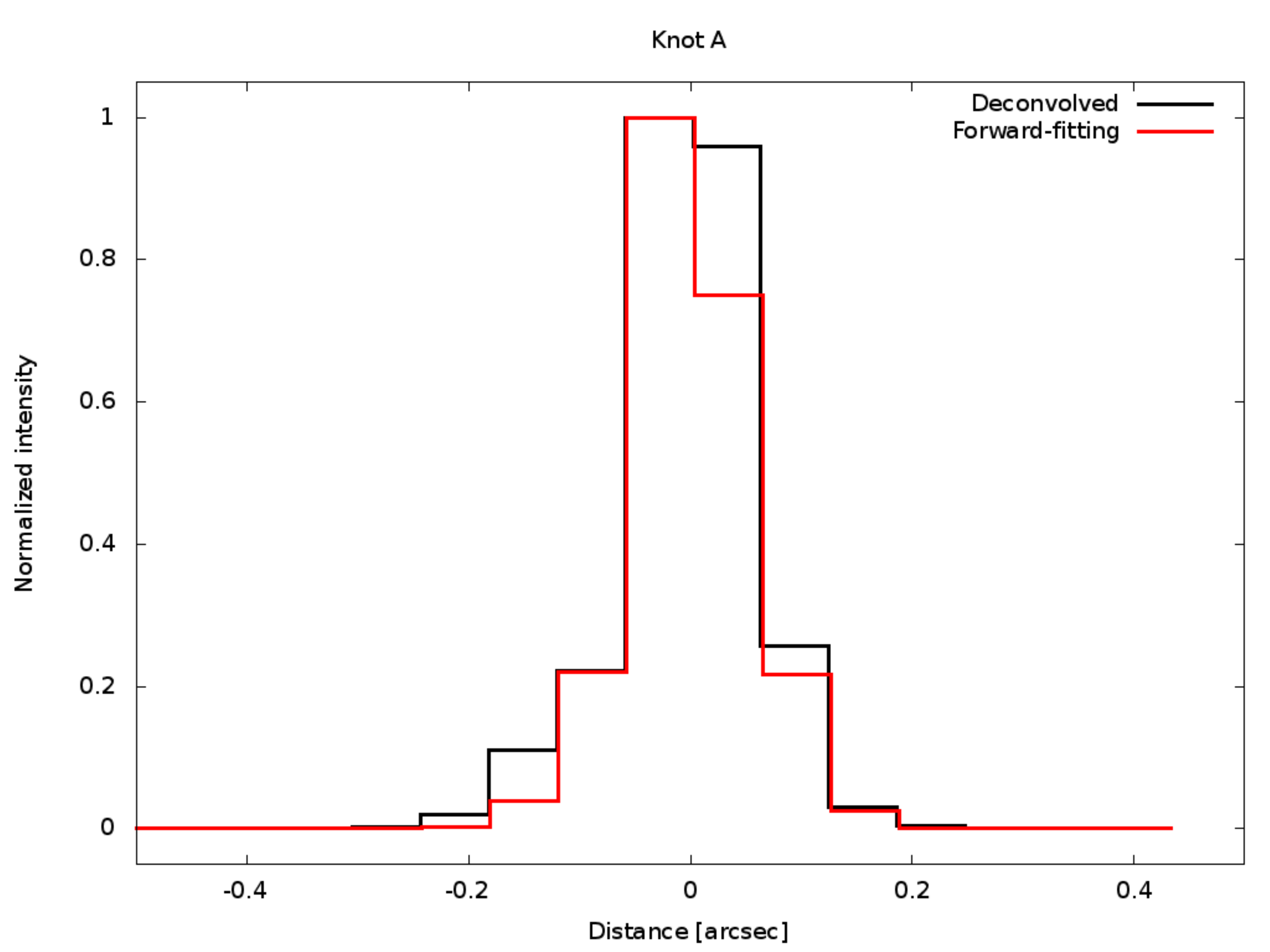}
    \includegraphics[width=0.33\textwidth]{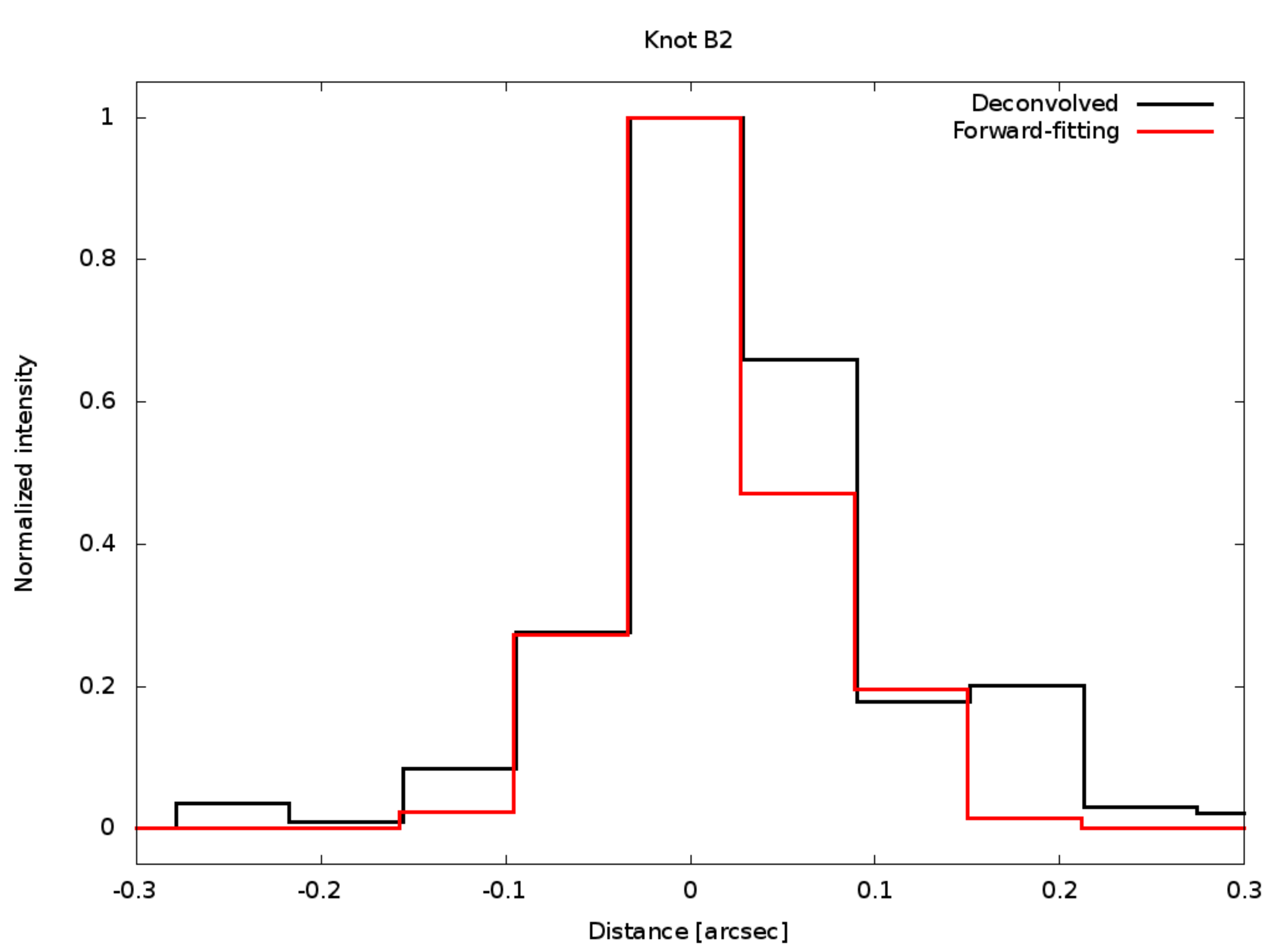}
    \caption{Transverse profiles of the brightest segments of the 3C\,273 jet (knots A, B2, C1, and H3) at different frequencies, obtained from the original {\it VLA} and the deconvolved {\it Hubble} and {\it Chandra} images. The last two panels present a comparison between the transverse X-ray profiles of knots A and B2 obtained by means of the deconvolution with the LRDA, and the multi-component forward-fitting.}
    \label{fig:trans_XUR}
\end{figure*}

The results of the analysis are given in Figure~\ref{fig:trans_XUR}. The lower panel in the figure presents in addition the direct comparison between the transverse X-ray profiles of knots A and B2 obtained with the two different methods of the image restoring, namely the deconvolution with the LRDA (\S\,\ref{S:deconvolution}), and the multi-component forward-fitting (\S\,\ref{S:forward}); these are in a good agreement. 

The quantitative evaluation of transverse sizes of knots by means of different methods in different wavelengths are summarized in Table~\ref{tab:trans}. 

Overall, the constructed profiles of the 3C\,273 jet reveal that the X-ray knots and their FUV counterparts have comparable transverse sizes, which are however smaller than the widths of the corresponding segments of the outflow measured at radio frequencies. It should be however noted that the radio profiles presented in Figure~\ref{fig:trans_XUR} are obtained from the original (i.e., not deconvolved) radio map; the radio knots' transverse sizes reported in Table~\ref{tab:trans} are instead estimated by fitting each knot with Gaussian using task {\tt jmfit} from AIPS software package.

\begin{table}[!b]
    {\scriptsize
        \begin{center}
            \caption{Transverse knot sizes in different wavelengths$^\dagger$.} 
            \label{tab:trans}
            \begin{tabular}{lllll}
                \hline\hline
                Knot & $w_X^{LRDA}$ & $w_X^{FF}$ & $w_U^{LRDA}$ &  $w_R^{JMFIT}$ \\
                \hline
                A   &  0.14   & $0.15 \pm 0.04$    &  0.13    &  $0.30 \pm 0.02$   \\
                B2	&  0.12   & $0.11 \pm 0.05$    &  0.18    &  $0.67 \pm 0.02$   \\
                C1	&  0.28   & $0.29 \pm 0.07$    &  0.28   &  $0.65 \pm 0.04$   \\
                H3	&  0.13   & $0.11 \pm 0.04$    &  0.18      &  $0.58 \pm 0.003$  \\
                \hline
            \end{tabular}
        \end{center}
        $^{\dagger}$All values are in arcsec. 
    }
\end{table}

\subsection{Longitudinal Jet Profiles}
\label{S:longitudinal}

\begin{figure*}[!t]
    \centering
    \includegraphics[width=\textwidth]{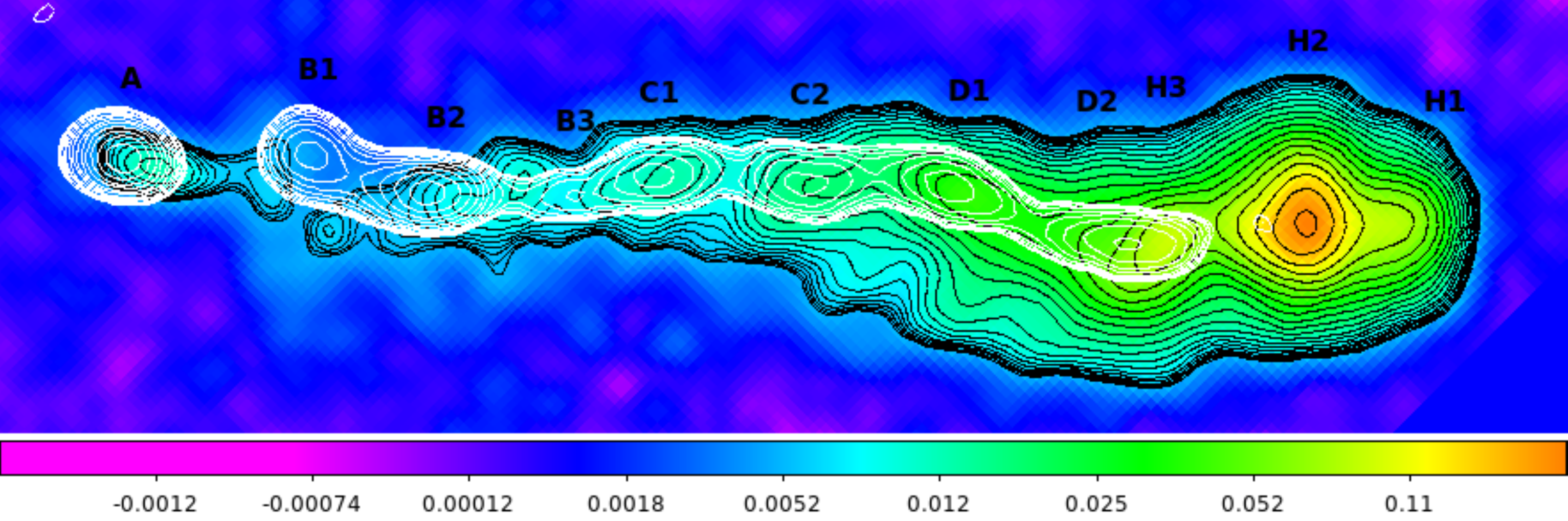}
    \caption{The radio image with radio (black) and the deconvolved X-ray (white) contours superimposed. The radio contour levels increase from 0.005 to 0.2 Jy/beam by a factor of $\sim 1.4$. The X-ray contour levels increase from 4 to 270 ``restored'' counts by a factor of $\sim 1.6$. The knots labeling following \citet{jester05}. The wiggling appearance of the outflow is clearly visible in the image.}
    \label{fig:XR_knots_con}
\end{figure*}

A visual inspection of Figure\,\ref{fig:XR_knots_con} uncovers a clear curvature of the 3C\,273 large-scale, followed by both the radio-- and the X-ray--emitting plasma; the FUV jet traces a similar, almost sinusoidal trail. In general, the X-ray and FUV knots can be matched with their prominent radio counterparts, although conspicuous differences between the X-ray/FUV and radio morphologies can be noted as well. First, as already emphasized in \S\,\ref{S:transverse}, the radio outflow is wider than the deconvolved X-ray/FUV jet. Second, the intensity peaks of the X-ray knots are found to be located systematically \emph{upstream} of the corresponding radio intensity peaks. In the case of the knot B1, this offset appears particularly dramatic, with the radio intensity peak displaced from the X-ray intensity peak not only in the longitudinal direction, but also in the transverse direction. Here we investigate such offsets more quantitatively. We note however that the exact positions of the FUV knots may be affected by the limited astrometry calibration related to the fact that the quasar core is not included in the analyzed {\it Hubble} field \citep[see the discussion in][]{jester07}; for this reason below we discuss only the {\it VLA} and {\it Chandra} data.

\begin{figure}[!b]
    \centering
    \includegraphics[width=0.45\textwidth]{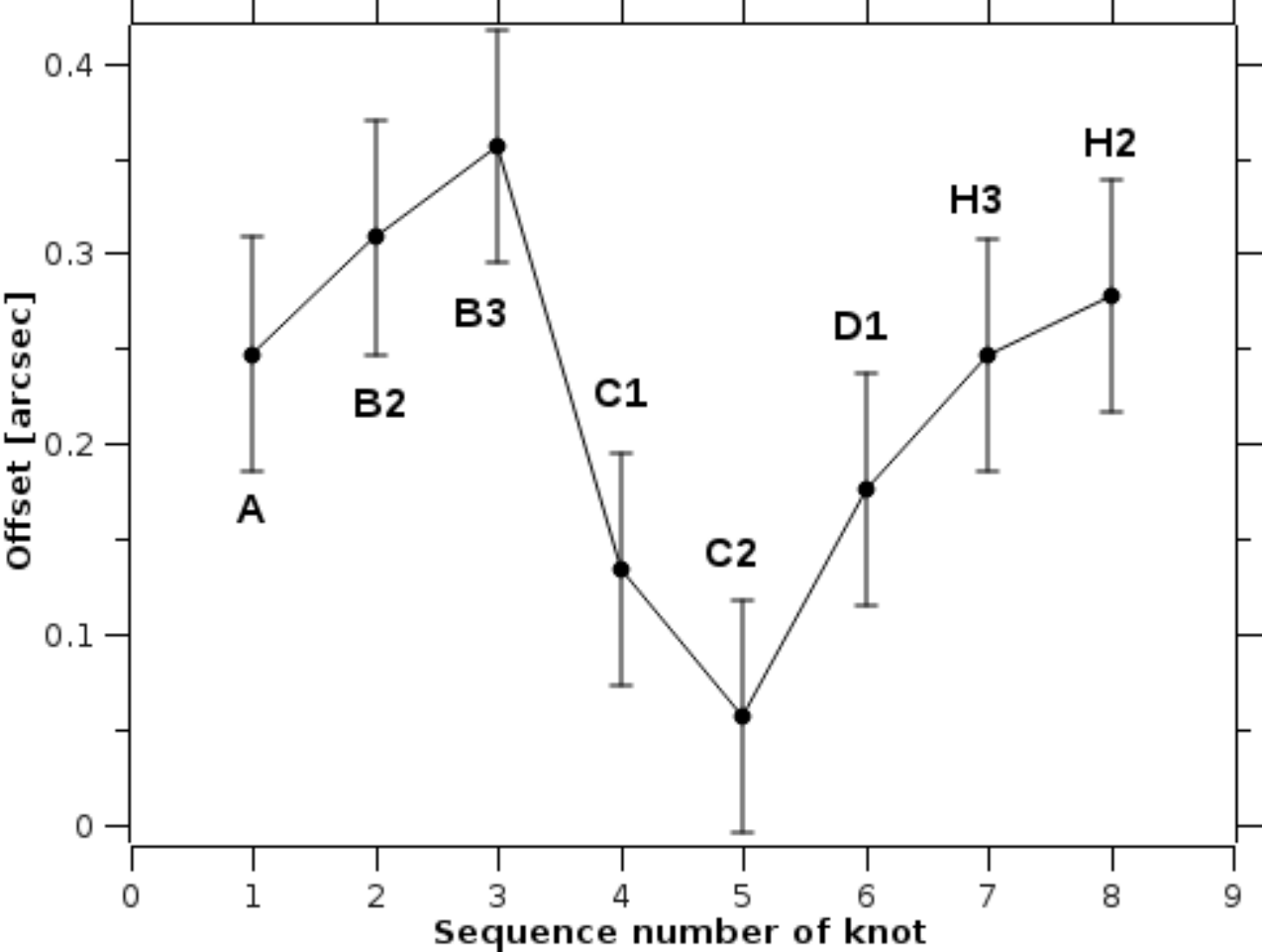}
    \caption{Spatial offsets between radio and X-ray intensity peaks in the 3C\,273 jet.}
    \label{fig:offset}
\end{figure}

Keeping in mind that the knots in 3C\,273 jet are not located along a straight line, we estimated the X-ray/radio positional offsets as
\begin{equation}
\Delta_{\rm X/R} =\sqrt{\left(x_{\rm X} - x_{\rm R}\right)^2 + \left(y_{\rm X} - y_{\rm R}\right)^2} \, ,
\end{equation}
where $(x_{\rm X}, y_{\rm X})$ and $(x_{\rm R}, y_{\rm R})$ are the exact coordinates of the intensity peaks of X-ray and radio knots, respectively, calculated using coordinates of the corresponding centroids on the deconvolved maps. We considered only those knots for which the peak positions can be measured with high accuracy; we also skip the problematic B1 segment of the outflow, in which case the identification of the X-ray knot with its radio counterpart is particularly confusing. The distribution of thus evaluated positional offsets, which range from $\simeq 0.05\arcsec$ up to $\simeq 0.35\arcsec$, is presented in Figure\,\ref{fig:offset}; the errors included in the figure correspond to the 1\,px\,=\,$0.0615\arcsec$ deviation. This distribution seems to reflect, at least to some extent, a wiggling appearance of the outflow depicted in Figure\,\ref{fig:XR_knots_con}. On the other hand, taking into account relatively small values of the offsets estimated above, which in fact are within the range of measurement uncertainties that can influence the estimate (namely, the {\it Chandra} astrometric accuracy, the PSF blurring, etc.), these results cannot be considered as a strong evidence, but instead as only a hint on a possible trend.

\section{Discussion}
\label{S:discussion}

As mentioned in the introduction, a number of observational findings regarding multiwavelength properties of large-scale quasar jets contradict the idea ascribing the observed X-ray jet emission to the inverse-Comptonization of the CMB photons by lower-energy electrons ($\gamma_e \sim 10^3$) characterized by a relativistic bulk motion (hereafter the `IC/CMB' model). 

First, the detections of X-ray counter-jets in high-power radio galaxies Pictor\,A and 3C\,353 (\citealt{hardcastle05,hardcastle16} and \citealt{kataoka08}, respectively), which are misaligned counterparts of radio-loud quasars, excludes any significant beaming involved. Second, prominent spatial offsets between the positions of radio and X-ray intensity peaks observed in many systems \citep[e.g.,][]{siemiginowska07,kataoka08}, along with the global longitudinal radio and X-ray jet intensity profiles \citep{hardcastle06,siemiginowska07}, as well as general constraints on the bulk jet velocities following from radio population studies \citep{mullin09}, are hard to be reconciled with the IC/CMB model, or at least with its `single-emission-zone' version. One possibility --- discussed in the particular context of 3C\,273 by \citet{uchiyama06}, and \citet{jester06,jester07} --- could be therefore that the low-energy (radio--to--infrared/optical) emission component is produced in the outer, slower jet boundary layer/jet cocoon, while the high-energy (optical/UV--to--X-ray) emission component is due to the IC/CMB process related to a fast jet spine. The optical/UV polarimetry of the PKS\,1136--135 jet, presented by \citet{cara13}, implies however that this high-energy continuum cannot be inverse-Compton in origin, but that it constitutes instead an additional, separate synchrotron component.\footnote{The optical polarimetry of 3C\,273 jet resulted so far in contradictory results: see \citet{roeser91} vs. \citet{thomson93}.} 

We note that the IC/CMB model was also claimed to be inconsistent with the detected X-ray variability of the Pictor\,A jet \citep{marshall10,hardcastle16}, and to stand in conflict with the $\gamma$-ray properties of the 3C\,273 system \citep{meyer14}. On the other hand, {\it Chandra} studies of distant radio-loud quasars seem to indicate that the IC/CMB emission may dominate radiative outputs of large-scale jets in, at least, high-redshift ``core-dominated'' systems, due to the strong increase of the CMB energy density in the early Universe, $\propto (1+z)^4$, as argued in \citet{cheung06,cheung08,cheung12}, and \citet{simionescu16}. Still, focusing the discussion on the particular case of the nearby 3C\,273, and keeping in mind all the observational results summarized above, from now on we continue our discussion assuming the `two synchrotron component' model for the broad-band jet emission.

The detailed analysis of the large-scale morphology of quasar 3C\,273 presented in this paper reveals that the FUV and X-ray jet emission is produced in a series of knots with transverse sizes of about $\simeq 0.5$\,kpc and lengths $\gtrsim 1$\,kpc, tracing the extremely well-collimated spine surrounded by a wider radio cocoon\footnote{But see in this context also \citet{marshall06}, who, based on the early {\it Chandra} dataset for the 3C\,273 jet, claimed ``unresolved knots that are smaller than the corresponding optically emitting knots and a broad channel that is about the same width as the optical interknot region''.}. The nature of the knots in quasar jets is in general subjected to the ongoing debate. In the case of hydrodynamical outflows with dynamically negligible magnetic field, such intensity enhancements may correspond to either stationary reconfinement shocks, or moving portions of the jet matter with excess kinetic power \citep[see the discussion in][]{stawarz04,marscher11,godfrey12}. The positional X-ray/radio intensity peak offsets, which, in the case of the 3C\,273 jet are hinted by the analysis presented here, could in principle be reconciled with the latter scenario \citep[see][]{kataoka08}.

On the other hand, in current-carrying jets, brightness enhancements may correspond to the development of large-scale magnetohydrodynamical instabilities, as observed in 3D simulations of relativistic, two-component outflows consisting of a central component with helical, dynamically relevant magnetic field, surrounded by a slower sheath/cocoon or a boundary layer with velocity shear \citep[see][and references therein]{mizuno07,mizuno11,mizuno14}. Interestingly, the ``wiggling'' appearance of the FUV/X-ray jet in 3C\,273, particularly prominent on the deconvolved maps, seems to be in tune with such a possibility. Moreover, a strong magnetization of the jet spine may in fact be consistent with the requirement for a continuous \emph{in-situ} acceleration of ultrarelativistic electrons along the outflow.

Radiative cooling timescales of synchrotron optical or especially X-ray emitting electrons in large-scale AGN jets are much shorter than the dynamical timescales involved. This implies the action of very efficient particle acceleration processes taking place along the outflows, which are not restricted solely to localized compact regions but instead distributed throughout the entire jet body \citep[see in this context the detailed analysis by][]{jester05,kataoka06}. It was proposed that these processes are related to stochastic interactions of radiating electrons with magnetic turbulence, expected to be particularly effective at the jet boundaries \citep[see, e.g.,][]{ostrowski00,stawarz02,rieger06,aloy08}. 

The other option, possibly more consistent with the multi-wavelength jet morphology of the 3C\,273 jet disclosed by our analysis, could be however magnetic reconnection taking place in a highly magnetized jet spine. It is interesting to note in this context the most recent results of the plasma kinetic simulations by \citet{sironi14}, \citet{guo14,guo16}, and \citet{werner16}, which revealed in accord that the reconnection process in relativistic plasma leads to the formation of power-law electron spectra, with energy indices $s_e \equiv - d \log N_e\!(\gamma_e)/d \log \gamma_e$ depending on the plasma magnetization parameter $\sigma$, and in particular increasing from $s_e \sim 1$ for $\sigma \gtrsim 10$ up to $s_e \gg 2$ for $\sigma \sim 1$. Such a dependance seems to be qualitatively consistent with the UV--to--X-ray spectral properties of 3C\,273 jet \citep[see][]{jester07}, assuming that the magnetization parameter within the jet spine --- traced by the chain of the FUV/X-ray brightness enhancements --- decreases gradually from the position of knot A up to the terminal hotspot H2/H3 \citep[projected distance from the quasar core $\approx 60$\,kpc, or de-projected $\sim 350$\,kpc for the expected jet viewing angle $\sim 10$\,deg; see][]{s04}.

On the other hand, it is not clear if the maximum electron energies available in the reconnection-related acceleration mechanisms captured in the aforementioned simulations are sufficient to account for the synchrotron emission extending up to X-ray frequencies, taking into account the expected sub-mG magnetic field intensity in large-scale quasar jets. Indeed, assuming a Poynting flux-dominated outflow with the total power $L_j$ not exceeding $10^{47}$\,erg\,s$^{-1}$, and predominantly toroidal magnetic configuration (for simplicity), the jet-frame magnetic field on large scales, $B' = \sqrt{4 L_j/c \, R_j^2 \Gamma_j^2}$, reads as
\begin{equation}
B' \sim 0.3 \left(\frac{L_j}{10^{47}\,{\rm erg/s}}\right)^{1/2} \left(\frac{R_j}{0.5\,{\rm kpc}}\right)^{-1} \left(\frac{\Gamma_j}{10}\right)^{-1} \, {\rm mG} \, ,
\end{equation}
where $R_j$ is the jet radius, and $\Gamma_j$ is the jet bulk Lorentz factor. With such, the production of synchrotron X-rays requires $\gamma_e \geq 10^7$ electrons.

\acknowledgments
V.\,M., M.\,O. and {\L}.\,S. were supported by the Polish National Science Centre through the grant DEC-2012/04/A/ST9/00083. We thank R. Perley for supplying the {\it VLA} map. We also acknowledge the anonymous referee, for her/his critical remarks and suggestions which helped to improve the paper.

\appendix
\section{The influence of PSF uncertainties on the source forward-fitting}
\label{sec:app1}

The image of a low-count point source (i.e., the image of a PSF in the case of a low photon statistics) can vary between different observations due to random photon fluctuations. Only for a number of counts large enough, the image approaches the expected shape defined by the {\it Chandra} PSF model for a given source position and energy spectrum. Therefore, the influence of photon fluctuations on the PSF shape determination, and hence on the source forward-fitting results, should be carefully recognized. 

Keeping in mind that in our paper we present as the final result the X-ray sizes of the brightest knots in the 3C\,273 jet, in the following analysis we used only the jet segment consisting of the first three bright knots (model components $G_1$, $G_2$ and $G_3$). For this, we have simulated 100 images of the PSF with the number of counts corresponding to the observational data, and different random seeds. Next, we created 100 models of the observational data, $M_i$, by convolving the simulated PSF images, $P_i$, with a source, $G^n$, given by the best PSF image ($P_0$; see Table~\ref{tab:4e3s}), 
\begin{equation}
M_i = \sum_{n=1}^N (P_i \circ G^n) + C \, ,
\end{equation}
where $C$ denotes the constant background component, and $N=3$ is the number of Gaussian components in the source model. These are therefore various ``observational'' realizations of the same source, and the differences between them are caused solely by the differences in the particular PSF images due to photon fluctuations.

\begin{figure}[p]
\centering
\includegraphics[width=0.49\textwidth]{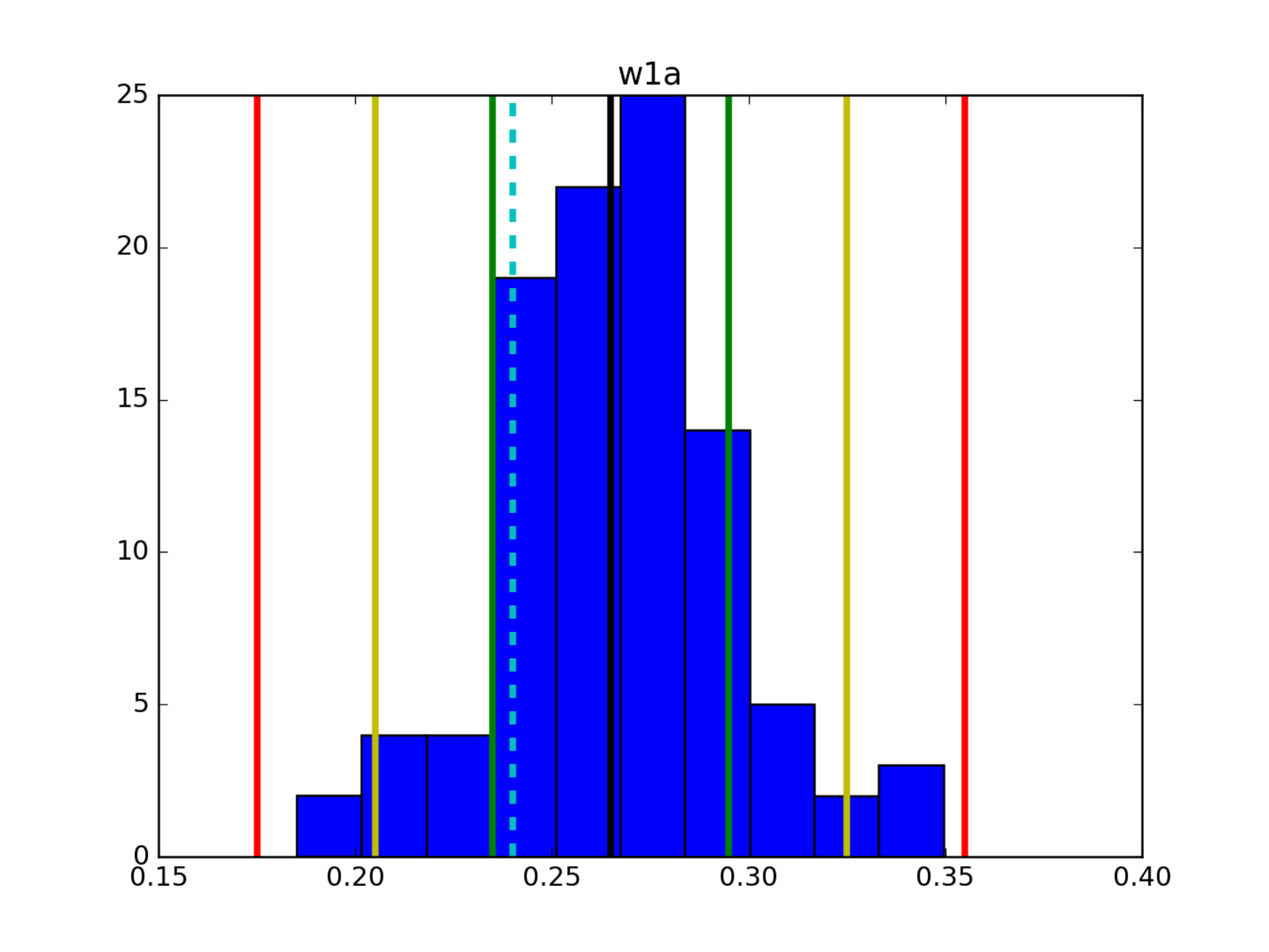}
\includegraphics[width=0.49\textwidth]{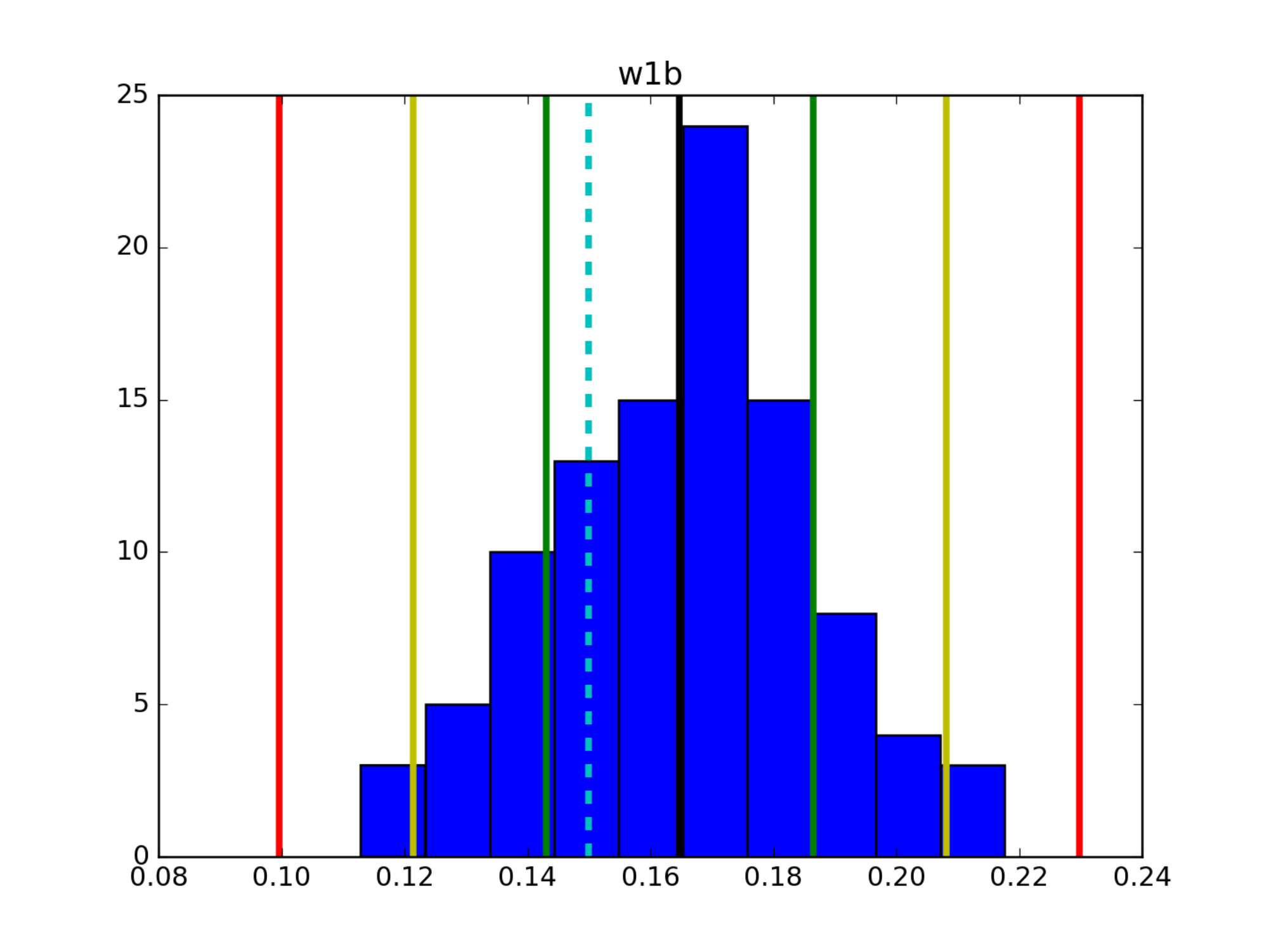}
\includegraphics[width=0.49\textwidth]{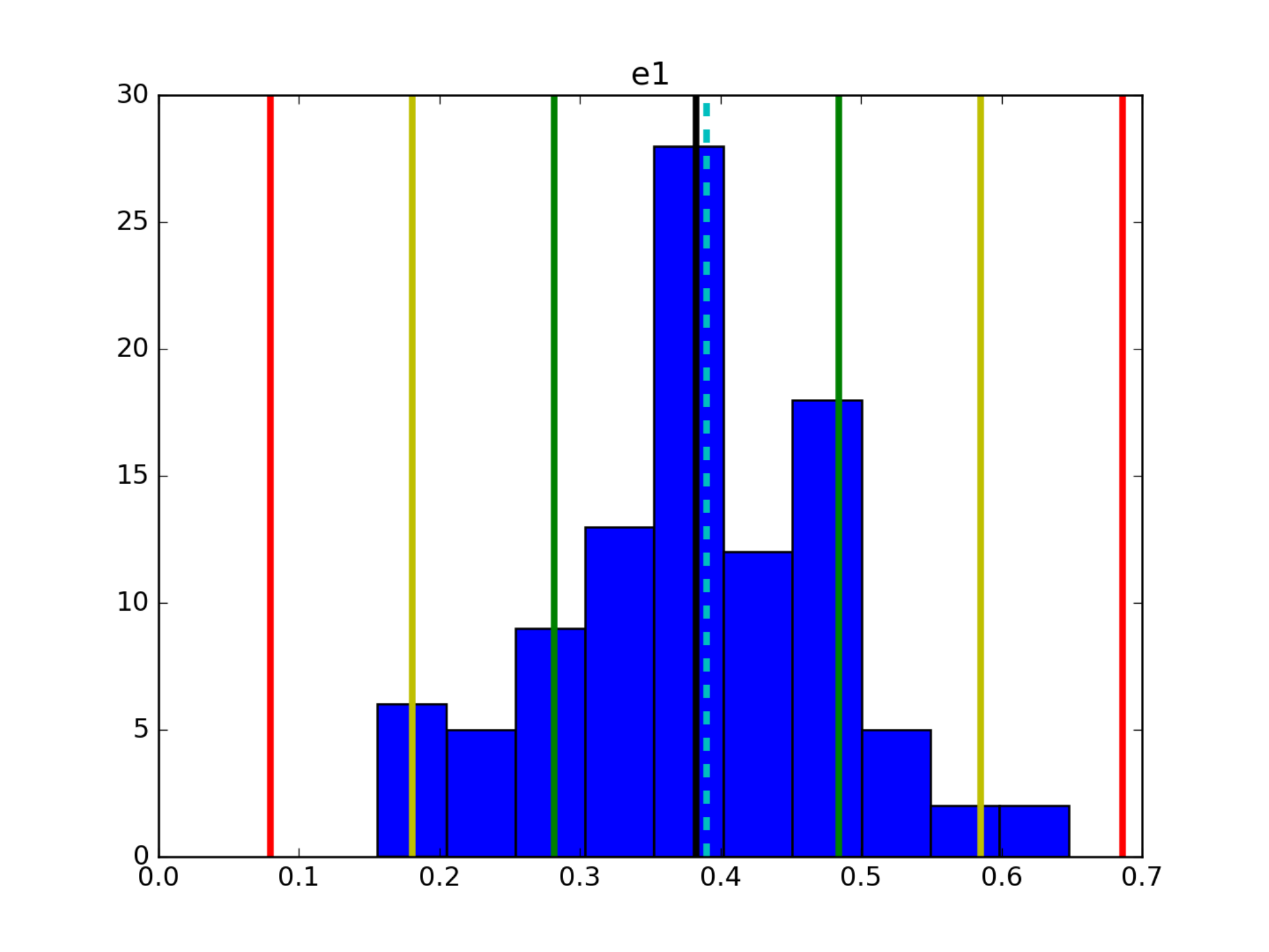}
\includegraphics[width=0.49\textwidth]{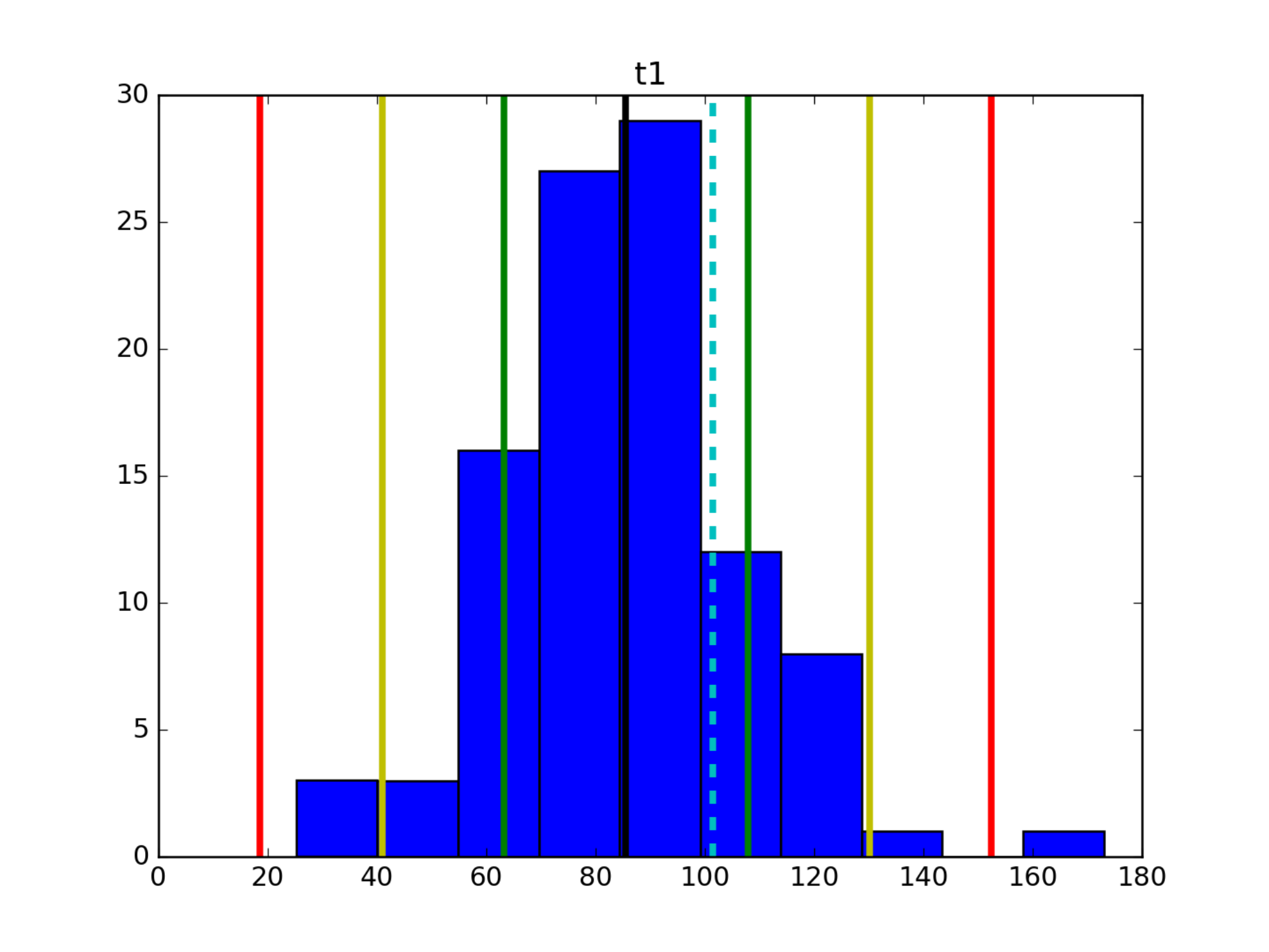}
\caption{The result of the forward-fitting for the source component $G_1$, using 100 models of the observational data. The component is modelled as an elliptical Gaussian with the major-axis FWHM $\mathrm{w1a}$, the minor-axis FWHM $\mathrm{w1b}$, the ellipticity $\mathrm{e1}$, and the position angle $\mathrm{t1}$. Vertical dashed cyan lines denote the parameter values calculated from the real data using the best PSF image $P_0$ (see Table\,\ref{tab:4e3s}). Vertical green, yellow, and red solid lines correspond to one, two, and three standard deviations in the respective simulated distributions.}
\label{fig:G1}
\end{figure}

\begin{figure}[p]
\centering
\includegraphics[width=0.49\textwidth]{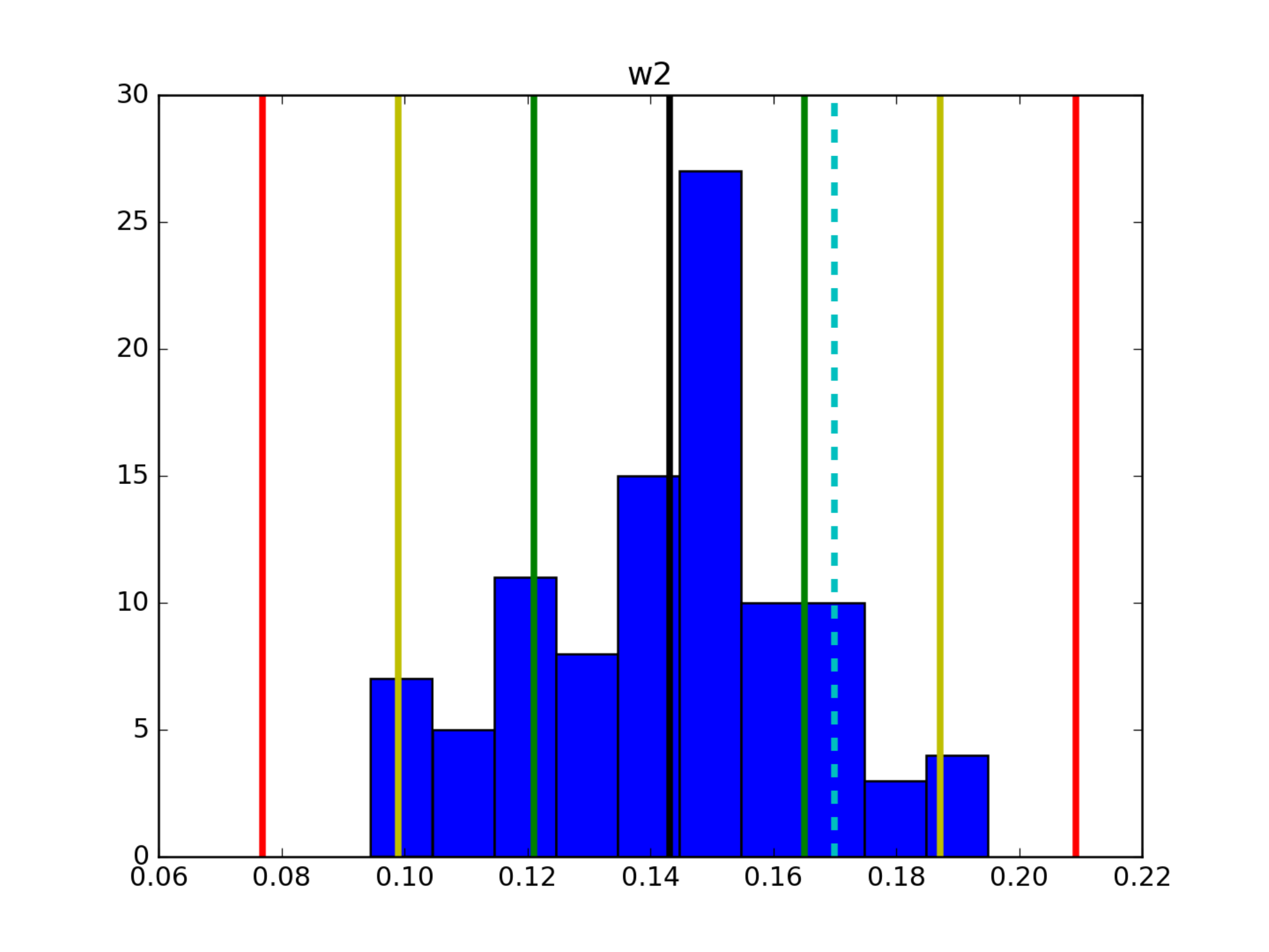}
\caption{Same as Figure~\ref{fig:G1}, but for the source component $G_2$, modelled as a spherical Gaussian with the FWHM $\mathrm{w2}$.}    
\label{fig:G2}
\end{figure}

\begin{figure}[p]
\centering
\includegraphics[width=0.49\textwidth]{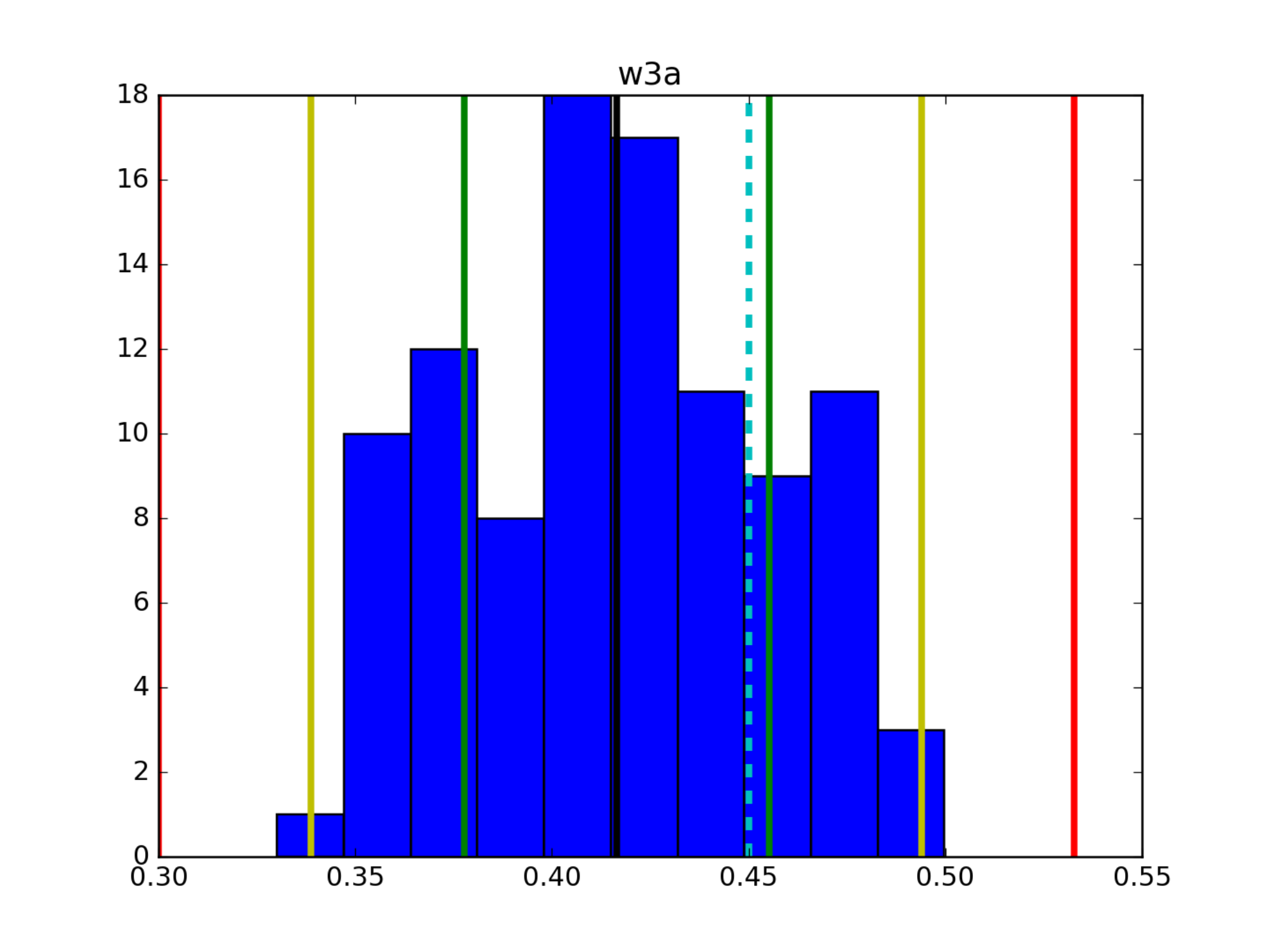}
\includegraphics[width=0.49\textwidth]{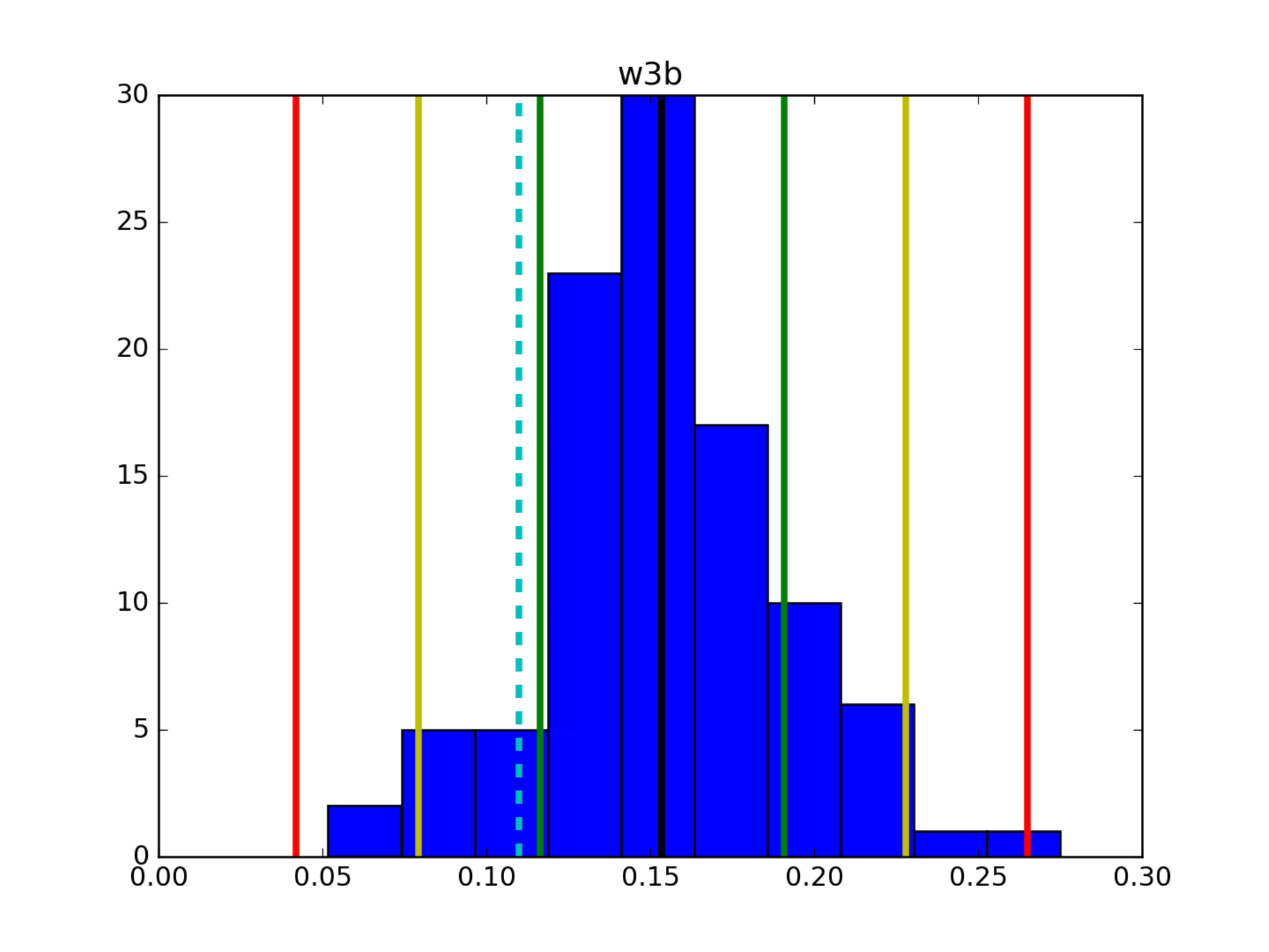}
\includegraphics[width=0.49\textwidth]{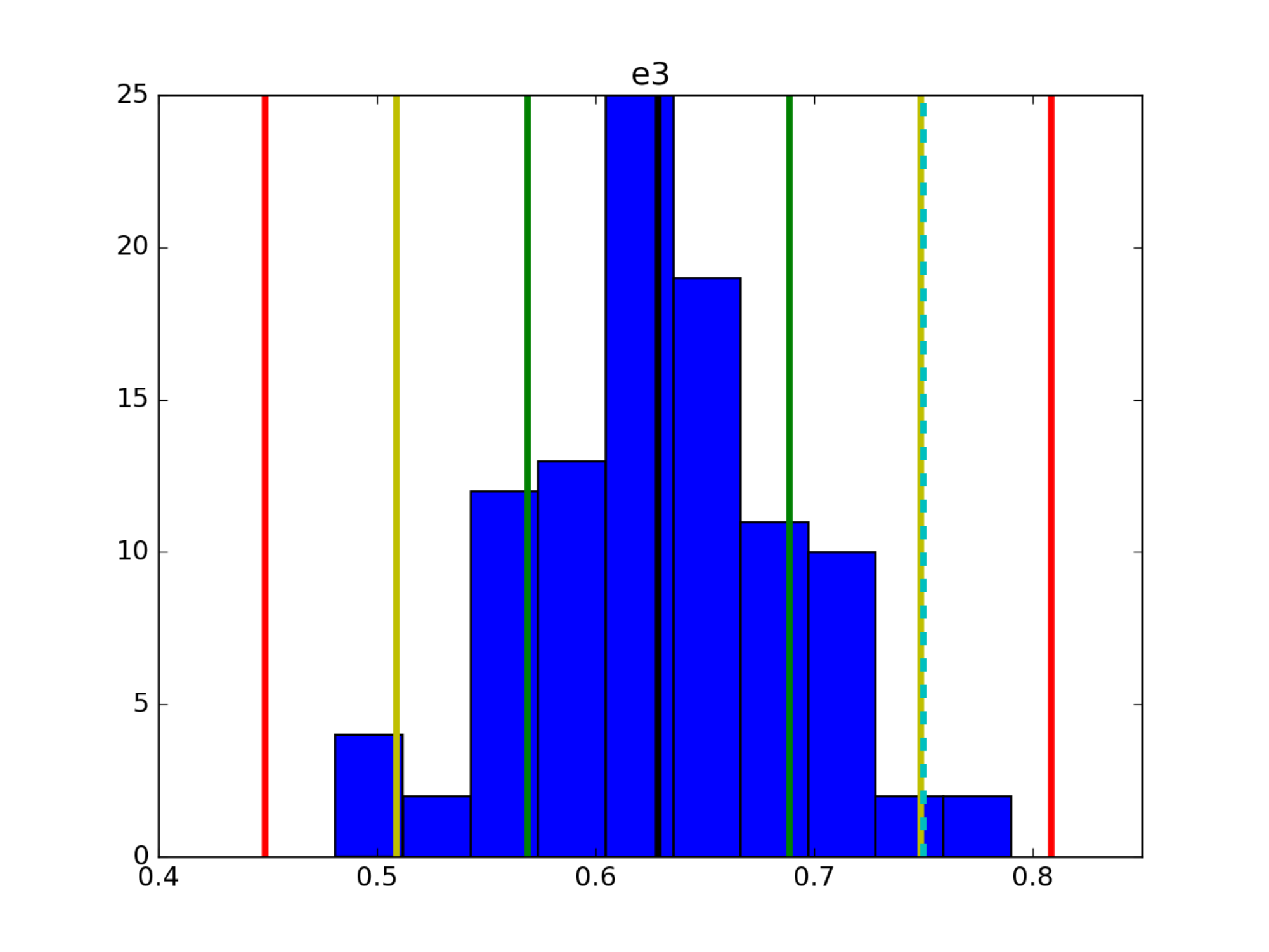}
\includegraphics[width=0.49\textwidth]{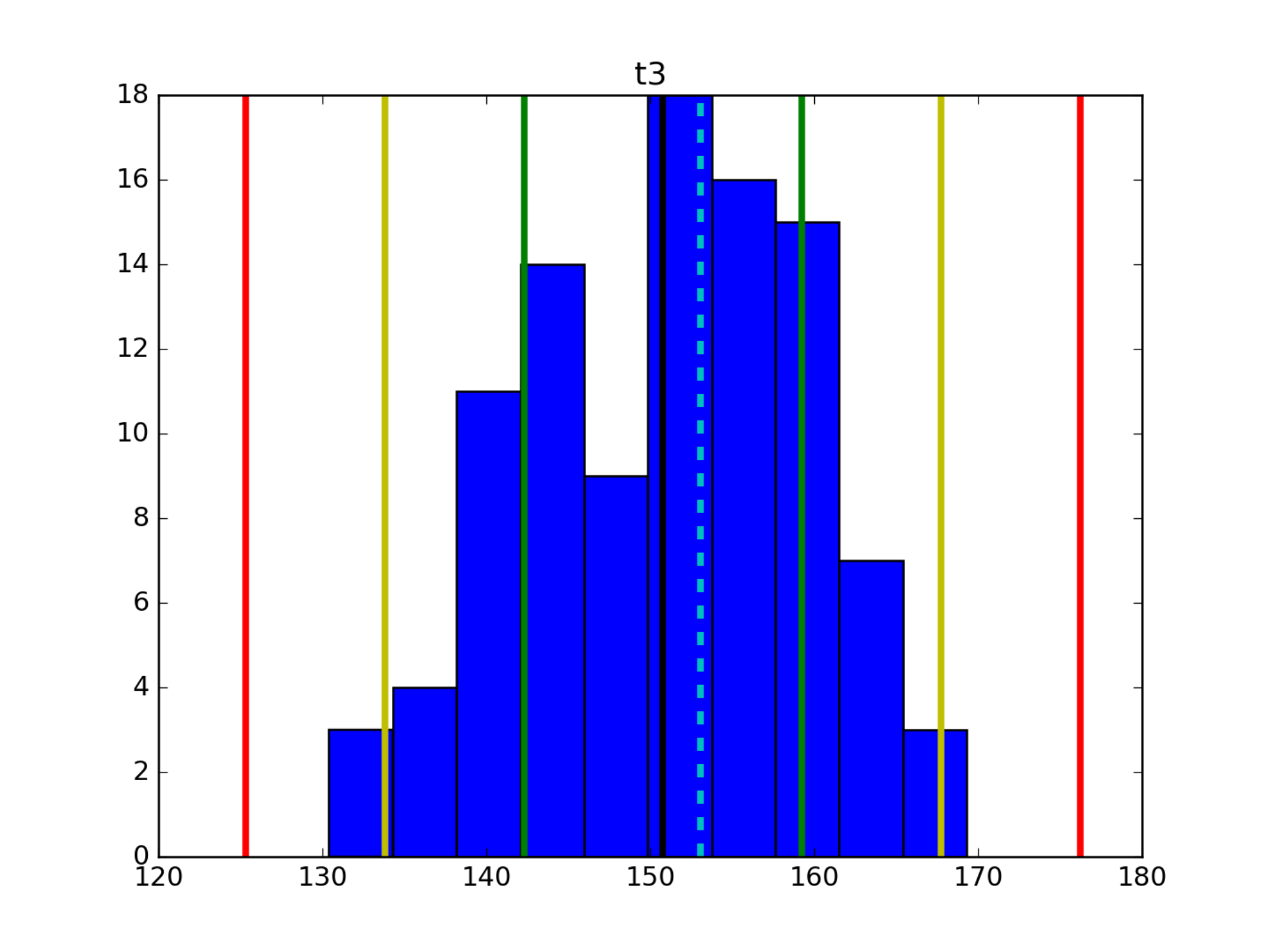}
\caption{Same as Figure~\ref{fig:G1}, but for the source component $G_3$, modelled as an elliptical Gaussian with the major-axis FWHM $\mathrm{w3a}$, the minor-axis FWHM $\mathrm{w3b}$, the ellipticity $\mathrm{e3}$, and the position angle $\mathrm{t3}$.}   
\label{fig:G3}
\end{figure}

We then performed the forward-fitting analysis for the models $M_i$ and the best PSF image $P_0$, just as in the case of the real data (see Section\,\ref{S:forward}), and analyzed the emerging scatter in the fitted source parameters. The results are presented in Figures~\ref{fig:G1}, \ref{fig:G2}, and \ref{fig:G3}. As shown, there is a reasonable agreement between the values of the source parameters calculated from the real data (denoted by vertical dashed lines in the figures) and the simulated data (histograms); moreover, most importantly, the range of three standard deviations in the simulated distributions are comparable with the errors on the corresponding parameters related to the forward-fitting procedure (Table\,\ref{tab:4e3s}).

\begin{figure}[p]
\centering
\includegraphics[width=0.49\textwidth]{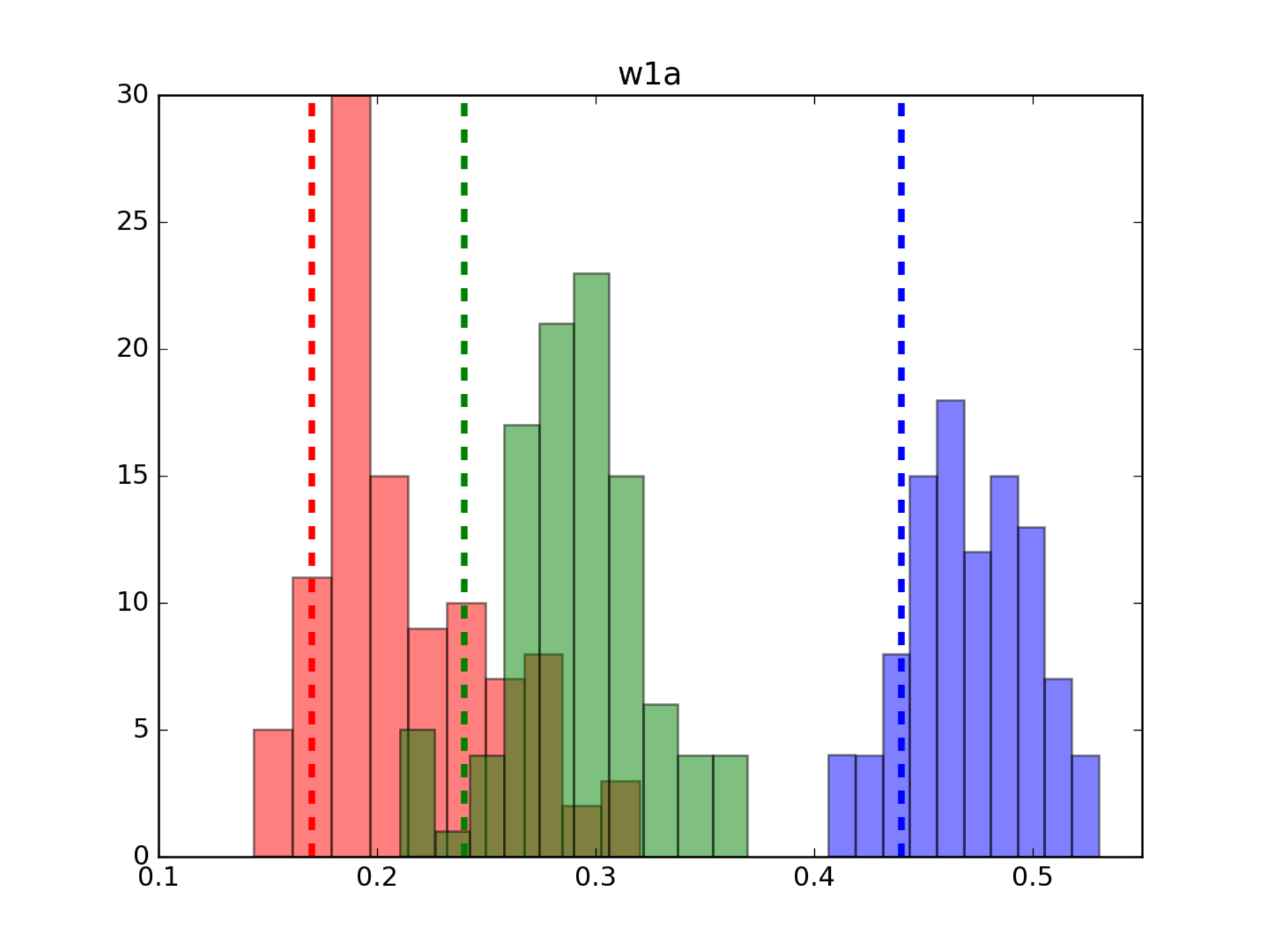}
\includegraphics[width=0.49\textwidth]{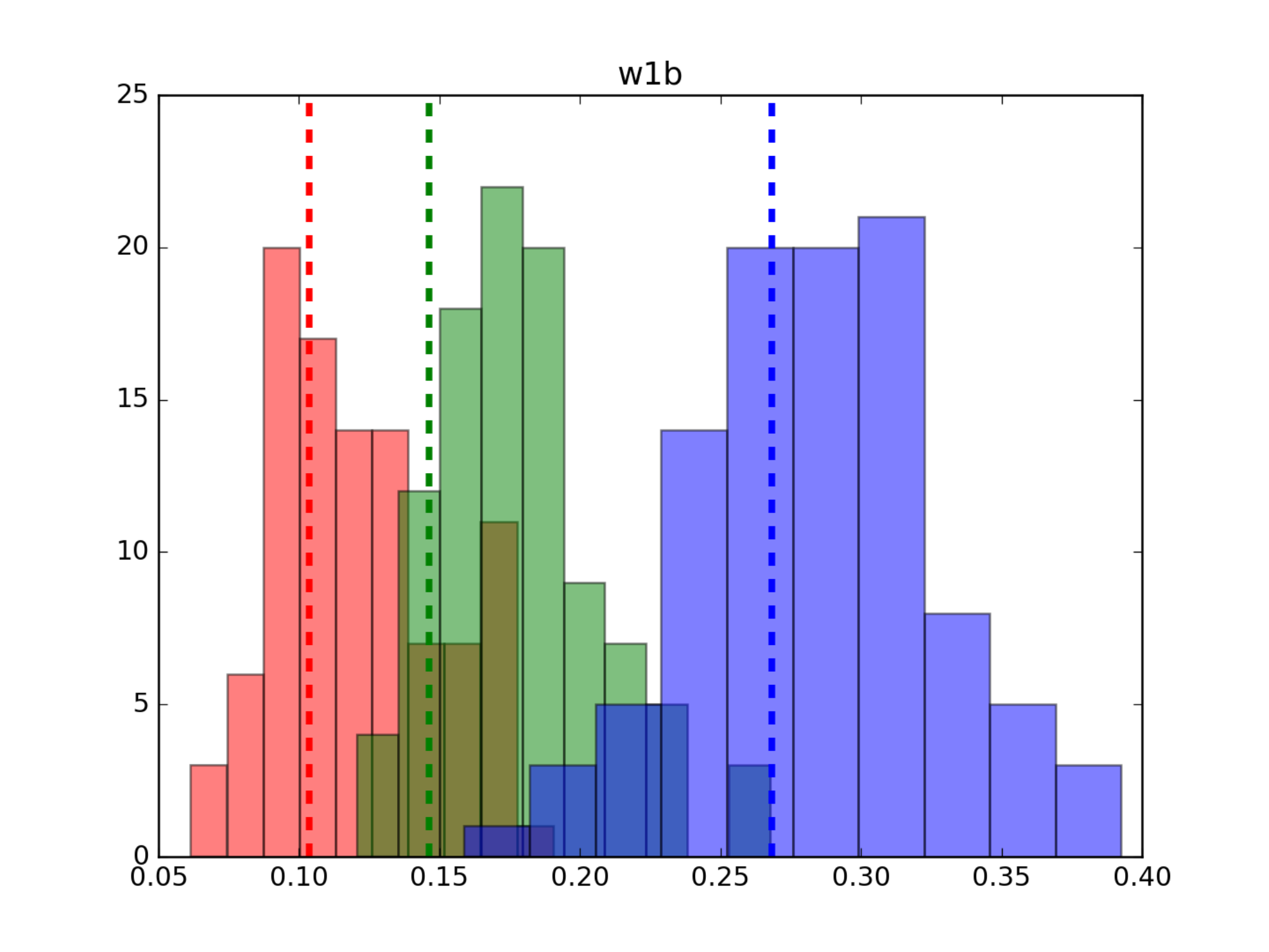}
\includegraphics[width=0.49\textwidth]{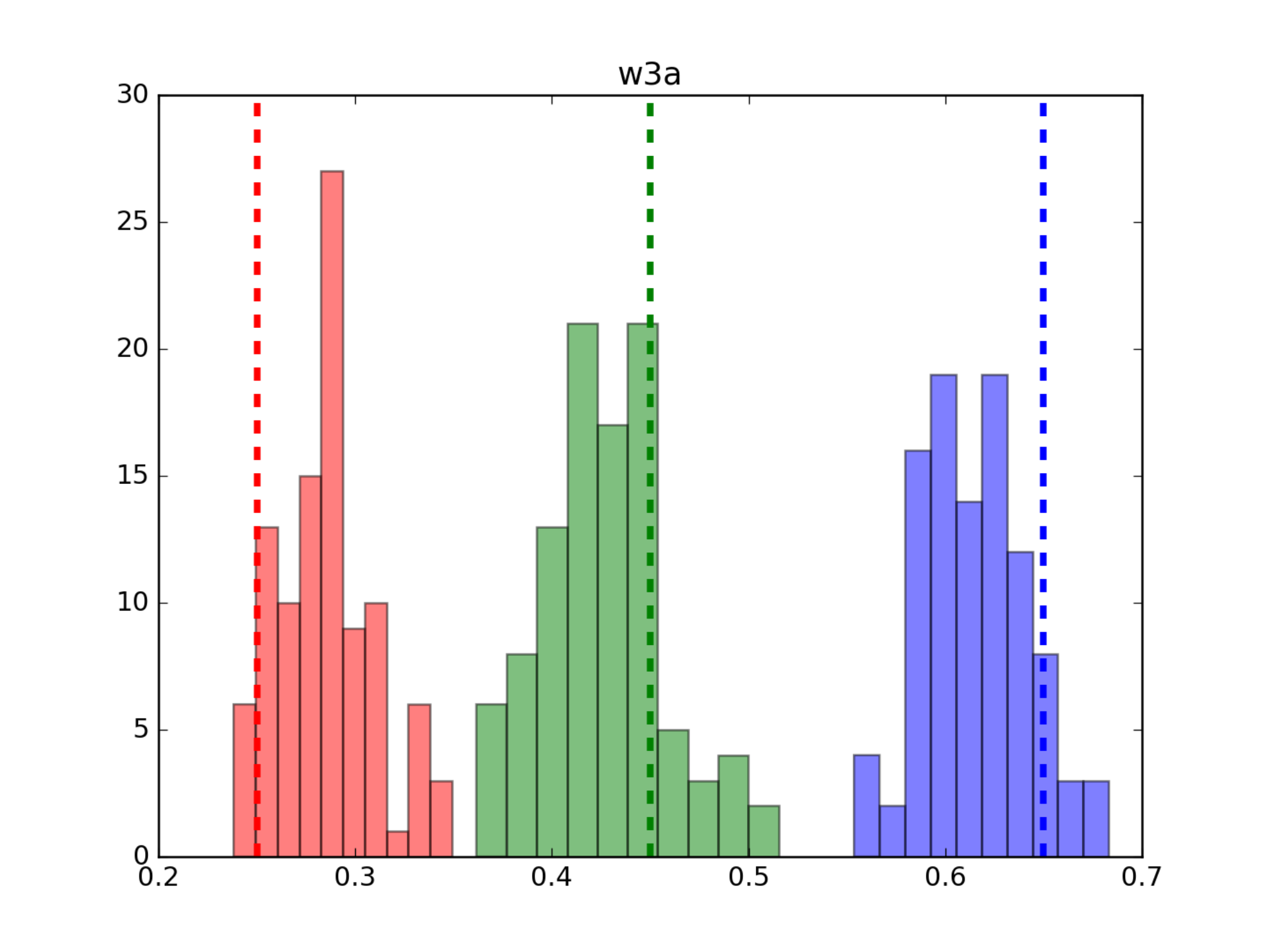}
\includegraphics[width=0.49\textwidth]{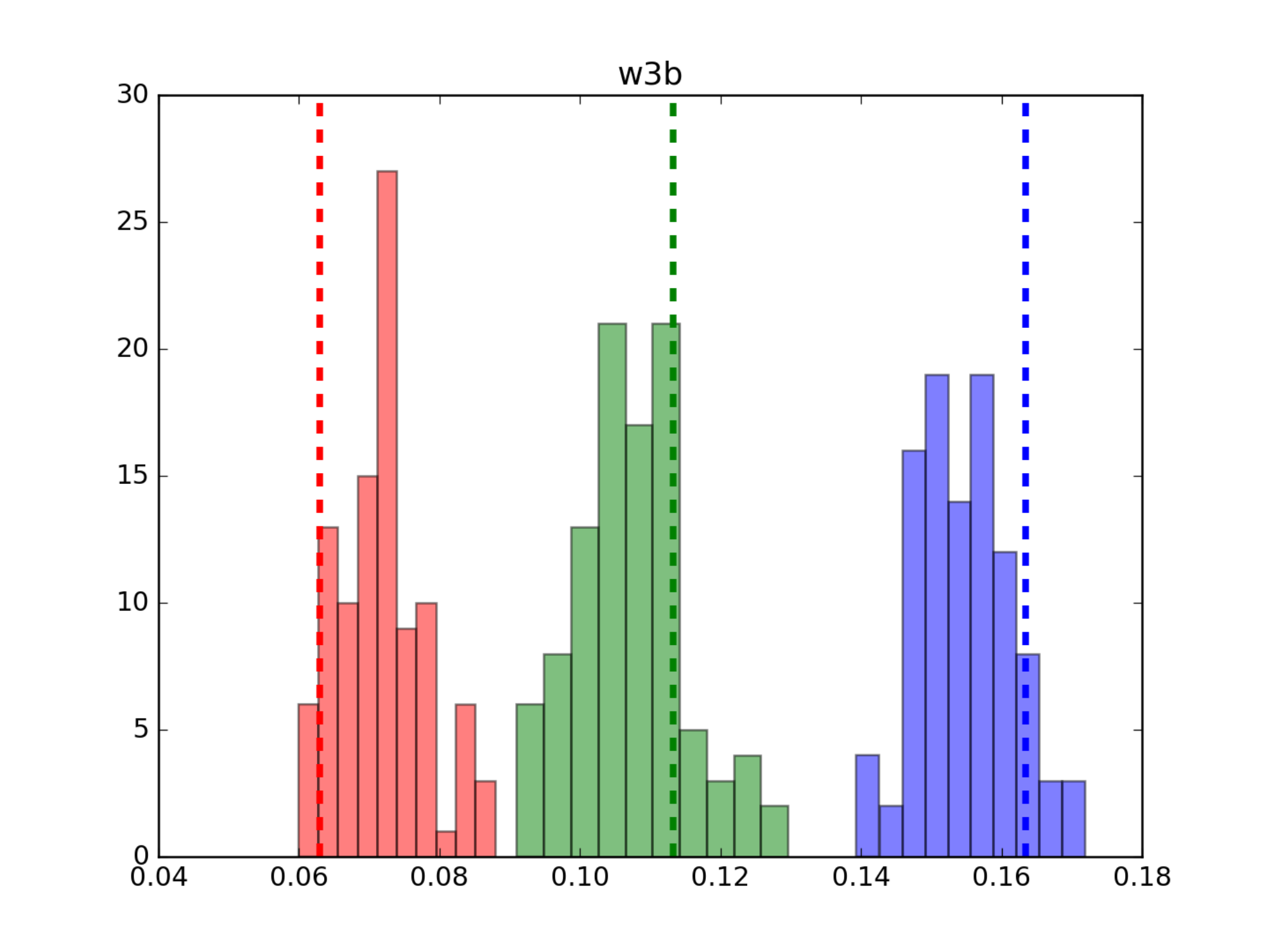}
\caption{The result of the forward-fitting for the source components $G_1$ (upper panels) and $G_3$ (lower panels), assuming different component sizes. The components are modelled as elliptical Gaussians with the major-axes FWHM $\mathrm{w1a}$, and the minor-axes FWHM $\mathrm{w1b}$. Vertical dashed lines denote the initial (assumed) FWHM values.}
\label{fig:G13fwhm}
\end{figure}

In order to investigate how the results of the source forward-fitting depend on the size of the selected source components, we have constructed three models with different sizes of the source components (i.e., different FWHM values). Taking into account that the component $G_2$ is not used in our final analysis, we have varied only the sizes of the components $G_1$ and $G_3$, and performed the simulations in an analogous way as described above. The results of the simulations are given in Figure~\ref{fig:G13fwhm}, where the initial values of the source FWHM are shown by dashed lines; as demonstrated, the forward-fitting procedure again adequately restores the source sizes, and the scatter observed between different simulations is again comparable with the errors of the forward-fitting procedure.

\section{Model selection in the source forward-fitting}
\label{sec:app2}

In order to select the best model for the source forward-fitting, we used the Akaike and Bayesian information criteria (AIC and BIC). For the six test cases $m_0,\dotsc,m_5$, where the index denotes the number of elliptical knots, the resulting AIC/BIC values are: $m_0:$\,39407.5/39654.0, $m_2:$\,39401.8/39665.9, $m_2:$\,39380.6/39662.3, $m_3:$\,39334.0/39633.3, $m_4:$\,39315.1/39631.9, $m_5:$\,39336.3/39670.8. For our analysis we have selected the case $m_4$ (four elliptical knots), characterised by the smallest AIC/BIC values. This, together with three additional spherical knots selected by visual inspection of the {\it Chandra} image, constituted the final model {\tt 4e3s} adopted in the source forward-fitting.

\begin{figure}[p]
    \centering
    \includegraphics[width=0.5\textwidth]{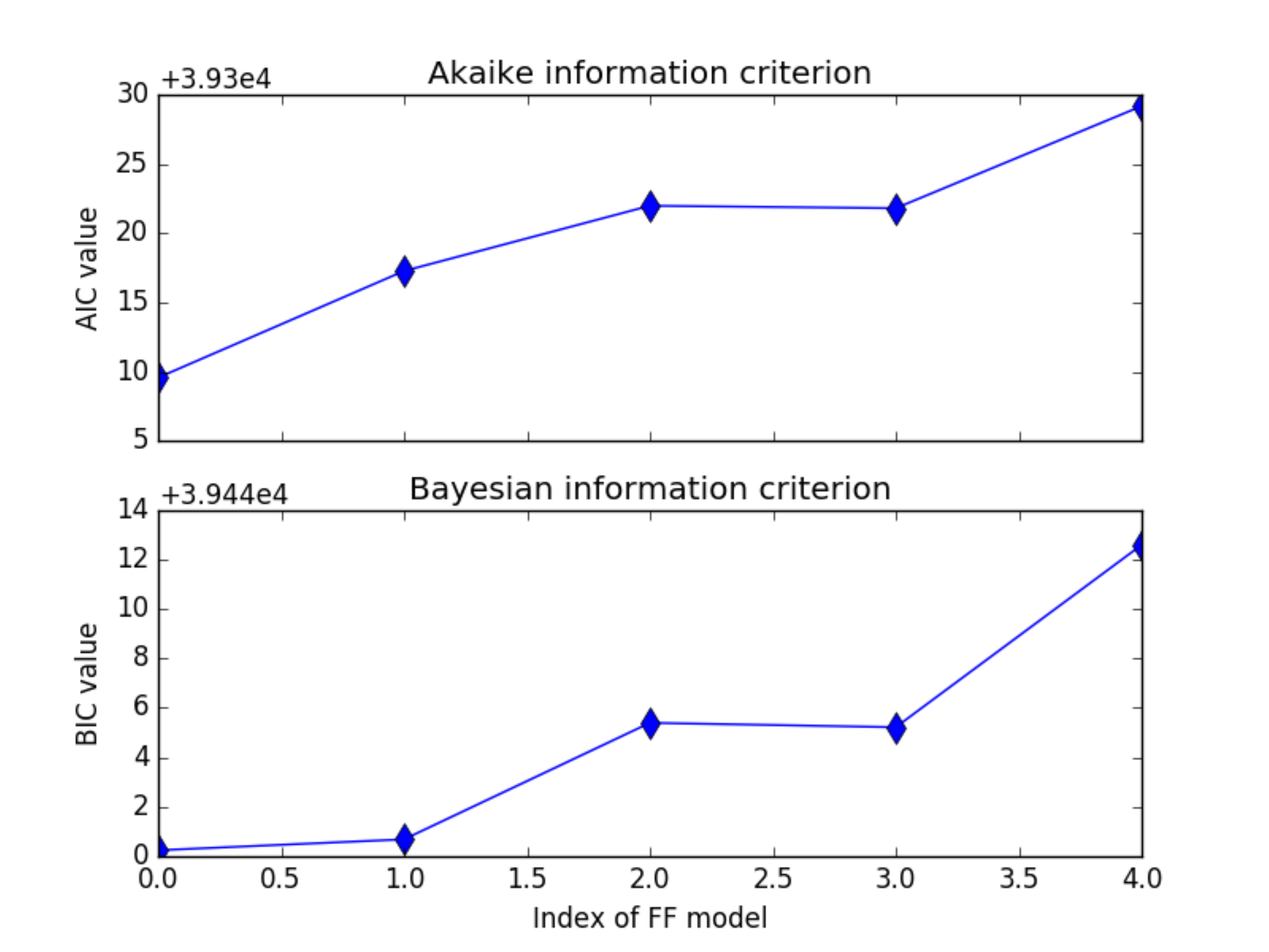}
    \caption{AIC/BIC values for different variants of the  {\tt 4e3s} model adopted in the source forward-fitting. The numbering on the horizontal axis corresponds to the variants of the model, and in particular ``0'' denotes the model {\tt 4e3s} with all the elliptical components extended, ``1'' denotes the model with the first elliptical model component reduced to a point-like source, ``2'' -- with the second elliptical component reduced to a point-like source, etc.}
    \label{fig:AIC-BIC}
\end{figure}

In order to confirm the extended nature of the knots in the framework of the selected model {\tt 4e3s}, we made the following analysis based on the information criteria: we have consecutively ``switched off'' extension of the model components, and calculated the corresponding AIC/BIC values for the model. Figure~\ref{fig:AIC-BIC} presents the results of the analysis; here, the numbering on the horizontal axis corresponds to the variants of the model considered, and in particular ``0'' denotes the model {\tt 4e3s} with all the elliptical components extended, ``1'' denotes the model with the first elliptical model component reduced to a point-like source, ``2'' -- with the second elliptical component reduced to a point-like source, etc. As shown in the plot, the original model {\tt 4e3s} is characterized by the minimum AIC/BIC values.

\end{document}